\documentclass[fleqn,usenatbib]{mnras}

\usepackage{newtxtext,newtxmath}
\usepackage{amsmath}
\usepackage[T1]{fontenc}

\DeclareRobustCommand{\VAN}[3]{#2}
\let\VANthebibliography\thebibliography
\def\thebibliography{\DeclareRobustCommand{\VAN}[3]{##3}\VANthebibliography}

\usepackage{graphicx}	
\usepackage{amsmath}	
\usepackage{soul}

\usepackage{lineno}

\newcommand{\kepler}[0]{\emph{Kepler}}
\newcommand{\bison}[0]{\emph{BiSON}}

\newcommand{\plato}[0]{\emph{PLATO}}
\newcommand{\teff}[0]{$T_{\text{eff}}$}
\newcommand{\rsolar}[0]{\rm R$_{\odot}$}
\newcommand{\msolar}[0]{\rm M$_{\odot}$}
\newcommand{\logg}[0]{$\log g$}
\newcommand{\yinit}[0]{$Y$}
\newcommand{\feh}[0]{{[Fe/H]}}
\newcommand{\mlt}[0]{$\alpha_{\rm MLT}$}
\newcommand{\smass}[0]{\rm M$_{\odot}$}

\newcommand{\Dnu}[0]{$\Delta\nu$}

\newcommand{\numax}[0]{$\nu_{\rm max}$}

\title[Systematics in Asteroseismic Modelling]{Systematics in Asteroseismic Modelling: Application of a Correlated Noise Model for Oscillation Frequencies}

\author[T. Li et al.]{
Tanda Li$^{1,2,3,4}$\thanks{E-mail: t.li.2@bham.ac.uk}, 
Guy R. Davies$^{3,4}$,
Martin Nielsen$^{3,4,5}$,
Margarida S. Cunha$^{6}$,
 \newauthor
Alexander J. Lyttle$^{3,4}$
\\
$^{1}$ Institute for Frontiers in Astronomy and Astrophysics, Beijing Normal University, Beijing 102206, China\\
$^{2}$ Department of Astronomy, Beijing Normal University, Beijing, 100875, China\\
$^{3}$ School of Physics and Astronomy, University of Birmingham, Birmingham, B15 2TT, United Kingdom\\
$^{4}$Stellar Astrophysics Centre (SAC), Department of Physics and Astronomy, Aarhus University, Ny Munkegade 120, DK-8000 Aarhus C, Denmark\\ $^{5}$Center for Space Science, NYUAD Institute, New York University Abu Dhabi, PO Box 129188, Abu Dhabi, United Arab Emirates\\
$^{6}$Instituto de Astrof\'\i sica e Ci\^encias do Espa\c co, Universidade do Porto, CAUP, Rua das Estrelas, PT4150-762 Porto, Portugal\\
}

\date{Accepted XXX. Received YYY; in original form ZZZ}

\pubyear{2020}

\begin{document}
\label{firstpage}
\pagerange{\pageref{firstpage}--\pageref{lastpage}}
\maketitle

\begin{abstract}
The detailed modelling of stellar oscillations is a powerful approach to characterising stars. However, poor treatment of systematics in theoretical models leads to misinterpretations of stars. Here we propose a more principled statistical treatment for the systematics to be applied to fitting individual mode frequencies with a typical stellar model grid. We introduce a correlated noise model based on a Gaussian Process (GP) kernel to describe the systematics given that mode frequency systematics are expected to be highly correlated. We show that tuning the GP kernel can reproduce general features of frequency variations for changing model input physics and fundamental parameters. Fits with the correlated noise model better recover stellar parameters than traditional methods which either ignore the systematics or treat them as uncorrelated noise. 
\end{abstract}





\section{Introduction}\label{sec:intro}


One-dimension stellar models have been widely used for decades to predict the structure, evolution, and oscillations of stars. Systematic errors are expected because stellar models contain assumptions and approximations (e.g., using the mixing-length theory to describe the convection) which do not perfectly reflect the actual physics in stars. In asteroseismic mode frequencies, the most well-known systematic error is the so-called surface term, which appears {as a frequency offsets} between observations and theoretical predictions computed with the best-fitting structural model. The surface term is caused by the poor modelling of near-surface layers of the star \citep[see details in][]{ball2017surface}, and it is a major source of error in theoretical mode frequencies \citep[$\sim$5$\mu$Hz at oscillation frequency with the largest amplitude, i.e., $\nu_{\rm max}$, for the Sun;][]{1996Sci...272.1286C}. { Treatment of the surface term normally follows a deterministic approach with parameterisation based on the so-called `surface correction' formulae  \citep[e.g.][]{2008ApJ...683L.175K,2014A&A...568A.123B,2015A&A...583A.112S}. Previous studies for the \kepler{} LEGACY sample \citep{2017ApJ...835..173S,2017ApJ...835..172L} showed that those correction formulae can give good fits to the frequency offsets \citep{2018MNRAS.479.4416C}. This surface correction is expected to be a smooth function of the mode frequency and some formulae also contain the mode inertia.}
Secondary systematic errors are caused by other missing physics. As an example, stellar magnetic activity, which is not included in most stellar codes, shifts mode frequencies to a noticeable degree. For the Sun, which is an aged and inactive G-type star, its frequencies {of low angular degree modes} ($\ell \leq$  3) shift up to 0.5$\mu$Hz during a solar cycle \citep{2007ApJ...659.1749C}. Recent findings based on \kepler{}  \citep{2009IAUS..253..289B} data have shown even larger frequency shifting (up to $\sim$2$\mu$Hz) in some sun-like stars \citep{2017A&A...598A..77K,2018A&A...611A..84S}. 
\citet{2017MNRAS.464.4777H} notes that the impact of magnetic activity can be treated as a part of the surface term. Parametrising the time variation of the surface correction could be a way to account for the activity-related frequency variation in solar-like oscillators. There are also model errors that have not been well studied or properly treated in modelling. For instance, \citet{2015MNRAS.447..680G} stated that systematic differences between observed and model frequencies would be expected {for} metal-poor stars if an incorrect $\alpha$-enhancement value is used in model computations. 

\begin{figure}
	\includegraphics[width=1.0\columnwidth]{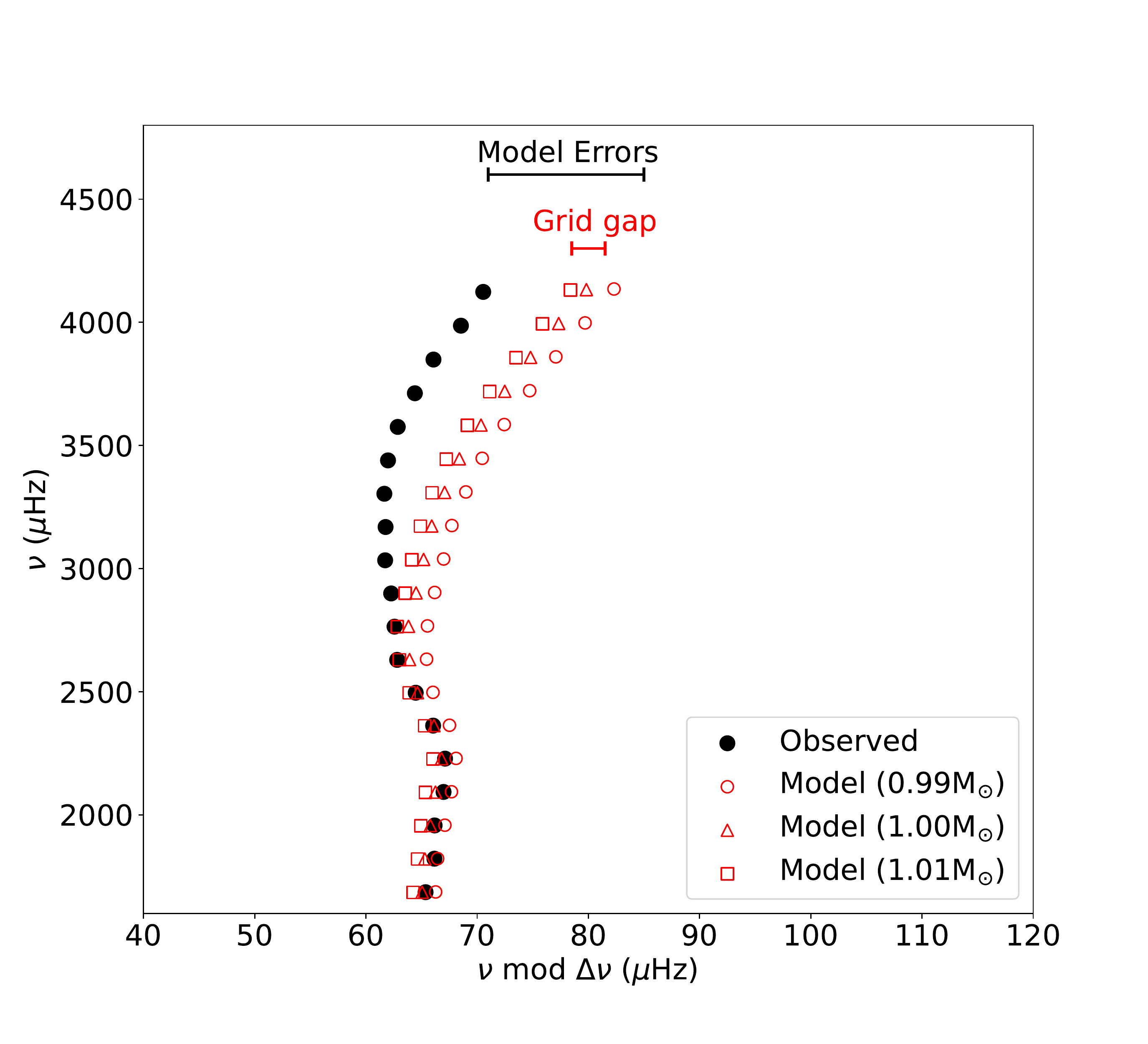}
    \caption{Systematic errors and uncertainties in theoretical mode frequencies. Black filled circles are radial ($\ell$ = 0) mode frequencies of the Sun observed by \bison\/ \citep{2017MNRAS.464.4777H} . Open symbols represent mode frequencies of three theoretical models with solar mean density but different input masses: $M$ = 0.99, 1.00, and 1.01\smass{}. Note that the models have the same input helium fraction (\yinit{} = 0.26), metallicity (\feh = 0.0), and mixing-length parameter (\mlt{} = 2.1). } 
  \label{fig:model_sys_ed}
\end{figure}

In grid-based modelling, systematic uncertainty could also undermine modelling solutions when the frequency resolution is comparable to the observed uncertainty. Here we define the frequency resolution as the difference between neighbouring points of the model grid. For a seismic model grid, neighbouring points are those with consecutive fundamental inputs (mass, matellicity, helium fraction, mixing-length parameter, etc.) and the same mean density. We use the mean density to locate neighbouring points because good-fitting models constrained by oscillation modes normally converge at a similar large separation (\Dnu{}), which tightly correlates to the mean density \citep[see][]{2022ApJ...927..167L}.
In Figure \ref{fig:model_sys_ed}, we demonstrate an example from sequential models of a model grid. Here we take the 1\smass{} model at the solar mean density, and we compare the model frequencies with its neighbouring grid points in terms of mass (0.99\smass{} and 1.01\smass{} models with the closest mean density). The mean frequency resolution is $\sim$1$\mu$Hz corresponding to a mass step of 0.01\smass. This frequency resolution is larger than the typical observed uncertainty of many well-studied solar-like oscillators\citep{2012A&A...543A..54A,2016MNRAS.456.2183D,2017ApJ...835..172L}. For instance, \citet{2020MNRAS.495.2363L} achieved an average uncertainty at about 0.2$\mu$Hz when measuring mode frequencies of 36 \kepler{} subgiants. This is to say, a sparse model grid could be significantly under-sampled for fitting to observed oscillation frequencies. 
%
%
%

Computing a fine model grid with a number of free model inputs is computationally expensive. Interpolating model frequencies based on established model grids like \textsc{AIMS} \citep[][]{2019MNRAS.484..771R} and \textsc{BASTA} \citep{2022MNRAS.509.4344A} or using a simplex method \citep[e.g. the `simplex search' function in \textsc{MESA Astero Module}][]{2015ApJS..220...15P} are possible solutions to this issue. These methods are statistically-sound but not efficient enough to apply to a large sample of stars. 
A fast approach is homologously scaling mode frequencies of a closely fitting model by a correction factor ($r$) to approximate a better fitting model \citep[][]{2008ApJ...683L.175K}. Scaling the mode frequencies by $r$ changes {the seismic large separation (\Dnu{})} by the same ratio and changes the mean density by a factor $r^2$. The downside of this method is that the $r$-scale transformation is not easily applied to other parameters, such as mass and age.

Traditional fitting strategies normally treat model systematics as white noise. For example, \citet{2020MNRAS.495.3431L} adopted a uniform white systematic noise when modelling the \kepler{} subgiants. The scale of noise is determined by the average frequency difference between the observation and the best-fitting model. This treatment is too simple to properly describe the model systematic uncertainty and hence leads to poorly measured uncertainties. 

The goal of this work is developing a better statistical treatment for systematics in model frequencies to improve the reliability of detailed modelling based on a stellar model grid. We propose a correlated noise model based on a Gaussian process (GP) kernel (also known as the covariance function) to describe the systematics. As a demonstration of the principle, we discuss the method with radial modes only, but it is extendable to all acoustic modes and also possibly to mixed modes. 
We introduce the underlying functions of the correlated noise model and discuss the fitting procedure in Section~\ref{sec:method}. We then apply the new fitting method to characterise fake model star in Section~\ref{sec:results} to examine whether fits with the correlated noise model better recover the {true stellar parameters}. Lastly, we close with some discussions about the new fitting method and conclusions of the paper in Section~\ref{sec:conclusion}.


\section{Method}\label{sec:method}


\subsection{Model Systematic Function}

%

\subsubsection{Understanding model systematics}
Understanding the systematics that affect the oscillation frequencies in a stellar grid is the key to finding proper functional forms with which to describe the model systematics. 
As seen in Figure \ref{fig:model_sys_ed}, the model systematics can be described as a combination of two components. The first is the frequency-dependent offset between the best-fitting model and the observed frequencies. Instead of calling it the surface term, we refer to it as the model systematic error ($\mathcal{E}$) to represent the systematics caused by all improper and missing physics in a theoretical stellar model. We use a smooth functional form, similar to a surface correction formulae, to model $\mathcal{E}$. 

\begin{figure}
\center
	\includegraphics[width=1.0\columnwidth]{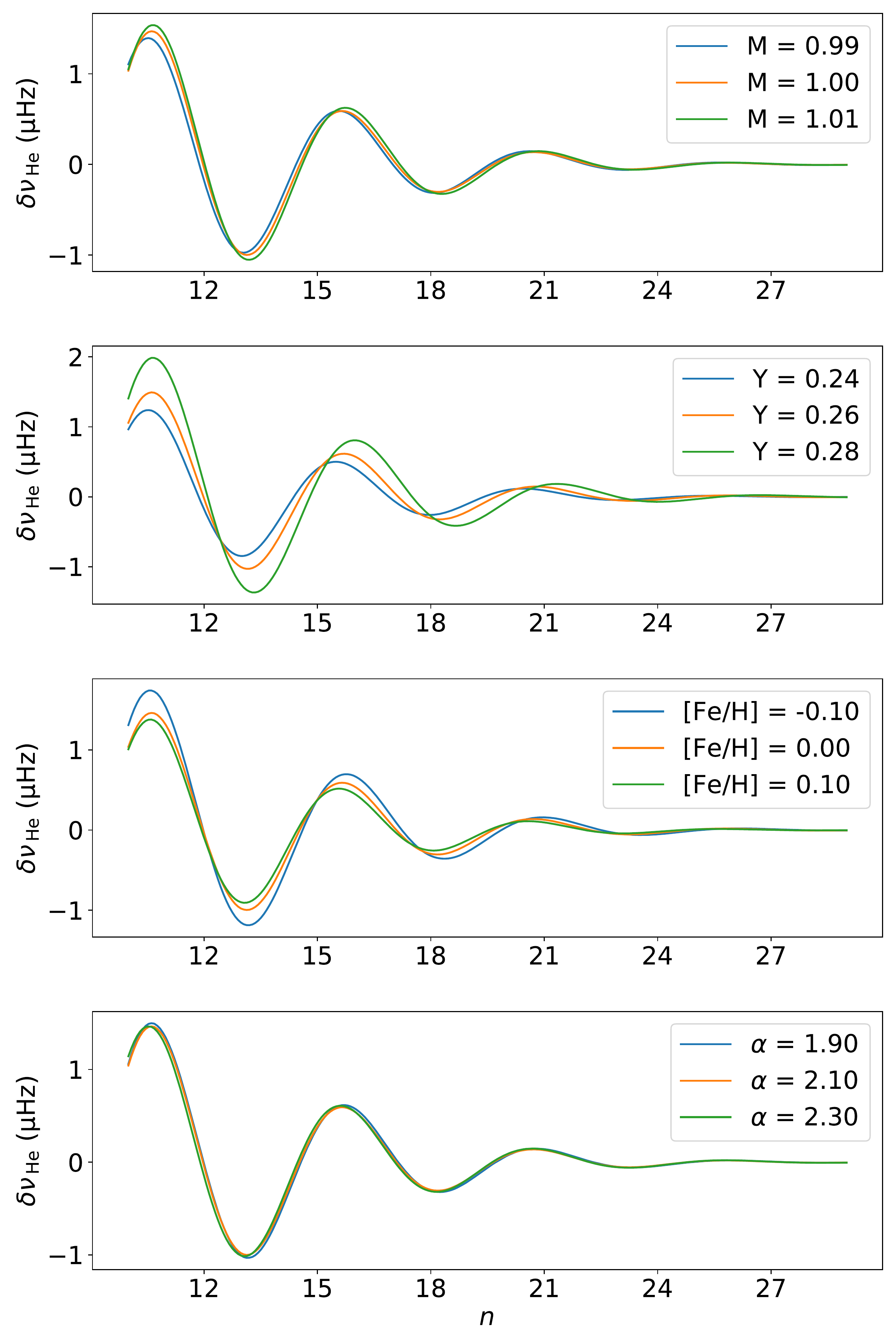}
    \caption{{The helium glitch features extracted with the tool \textsc{Asterion} (\url{https://github.com/alexlyttle/asterion}) from theoretical radial mode frequencies. 1st: models with different mass ($M$ = 0.99, 1.00, and 1.01${\rm M_{\odot}}$), but same metallicity (\feh{}= 0.0), helium fraction (\yinit{}= 0.26), and mixing-length parameters (\mlt{} = 1.9); 2nd: models with same mass ($M$ = 1.00${\rm M_{\odot}}$), metallicity (\feh{}= 0.0), and mixing-length parameters (\mlt{} = 1.9) but different input helium fractions (\yinit{}= 0.28, 0.26, and 0.24); 
 3rd: models with same mass ($M$ = 1.00${\rm M_{\odot}}$), helium fraction(\yinit{}= 0.26), and mixing-length parameters (\mlt{} = 1.9) but different metallicity(\feh{}= -0.1, 0.0, +0.1); 4th: models with same mass ($M$ = 1.00${\rm M_{\odot}}$), metallicity(\feh{}= 0.0), and helium fraction (\yinit{} = 0.26) but different mixing-length parameters (\mlt{}= 1.9, 2.1 and 2.3). All models have approximate solar mean density. }} 
  \label{fig:model_sys_glitch}
\end{figure}

The second component, as mentioned in Section~\ref{sec:intro}, is the systematic uncertainty ($\mathcal{U}$) caused by the frequency resolution of the model grid. 
The detailed modelling uses global parameters (effective temperature, luminosity, metallicity, etc.) and individual mode frequencies to characterise stars.
Comparing the models in Figure \ref{fig:model_sys_ed} we observe that, at a given mean density, changing the mass by a typical mass interval in a grid shifts the mode frequencies horizontally by an amount which increases smoothly with frequency.  Changing one of the other model inputs like metallicity, helium fraction, and mixing-length parameter shifts mode frequencies in a similar way. 
There is also a secondary term to be considered in the systematic uncertainty related to the signature of rapid structural variations in the oscillation frequencies (known as the helium `glitch' signature). 
The structural variation can be seen in the first adiabatic index ($\Gamma _{1}$). \citet{1990LNP...367.....O} and \citet{2007MNRAS.375..861H} assumed that the helium glitch signature arises from the second helium ionisation zone. Later studies indicate that this signature is from the $\Gamma_1$ peak between the first and second helium ionisation zones \citep[see][for details on modelling of the ionisation region]{2021A&A...655A..85H}. The helium glitch strongly correlates with the helium fraction in the convective envelope.
In Figure \ref{fig:model_sys_glitch}, we illustrate the signature of the helium glitches extracted from theoretical models with approximately the solar mean density. The glitch signatures follow the sinusoidal function with decaying amplitudes and the signature parameters (amplitude, period, and phase) change with the fundamental inputs \citep[][]{2007MNRAS.375..861H, 2019MNRAS.483.4678V}. For the models presented here, the average frequency shift is at a level of $\sim$0.1$\mu$Hz, which are comparable or larger than the observed frequency uncertainties on some \kepler{} stars \citep[e.g.,][]{2020MNRAS.495.2363L} thus should not be ignored.
Hence, the systematic uncertainty can be described with a two-term functional form. The primary term ($\mathcal{U}_{1}$) which is a very smooth function of frequency and the secondary term ($\mathcal{U}_{2}$)  which is a fast-varying function of frequency.

As a result, we describe the model systematics as the combination of the systematic errors ($\mathcal{E}$) and two systematic uncertainty terms ($\mathcal{U}_{1}$ and $\mathcal{U}_{2}$).

\subsubsection{The Application of a GP Kernel}

As demonstrated above, proper descriptions of systematic errors and uncertainties are smooth functions of frequency. This is expected because mode frequencies are highly correlated following the so-called asymptotic relation. Thus, a white noise model for the systematic noise term is inherently not a good model. Here we suggest a noise model with a smooth function form which is able to consider the correlation between mode frequencies, and we refer to it as correlated noise model (CNM). 

{The Gaussian Process (GP) kernel, which is used to generate a covariance matrix for a multivariate Normal distribution, is particularly suitable for building up the CNM.
\citet{2006gpml.book.....R} consider a GP in a functional form domain whereby the conditional GP acts to marginalize over all possible functional forms weighted by the prior (i.e. the kernel) and the data.  
Here we chose a Squared Exponential (SE) kernel as it follows a smooth functional form and is able to consider the correlation between oscillation frequencies. The SE kernel is described by the covariance matrix}

\begin{equation}\label{eq:sek2}
k(\nu, \nu ')  =  \sigma^2 {\rm exp}\left( - \frac{(\nu - \nu ')^2}{2l^2}\right). \\
\end{equation}
The kernel $k(\nu, \nu ')$ models the joint variability of the Gaussian Process random variables. {It returns the modelled covariance between {each pair of frequencies} $\nu$ and $\nu'$. When using the kernel to represent the systematic uncertainties, $\nu$ are the computed mode frequencies in the model grid and $\nu'$ {are the predicted frequencies based on $\nu$}.}
There are two free parameters in the function: the lengthscale $l$ and the noise variance $\sigma$. The lengthscale describes how smooth a generated function is (length of the `wiggles'), and the noise variance specifies the average distance of the function away from its mean. {Tuning these two parameters allows us to generate functional noise forms with different smoothness and varying ranges. In sampling process, the kernel randomly generates frequency noise ($\nu$ - $\nu'$) for a given $\nu$ following the multivariate normal distribution specified by the lengthscale and the noise variance.} 

\begin{figure}
	\includegraphics[width=1.0\columnwidth]{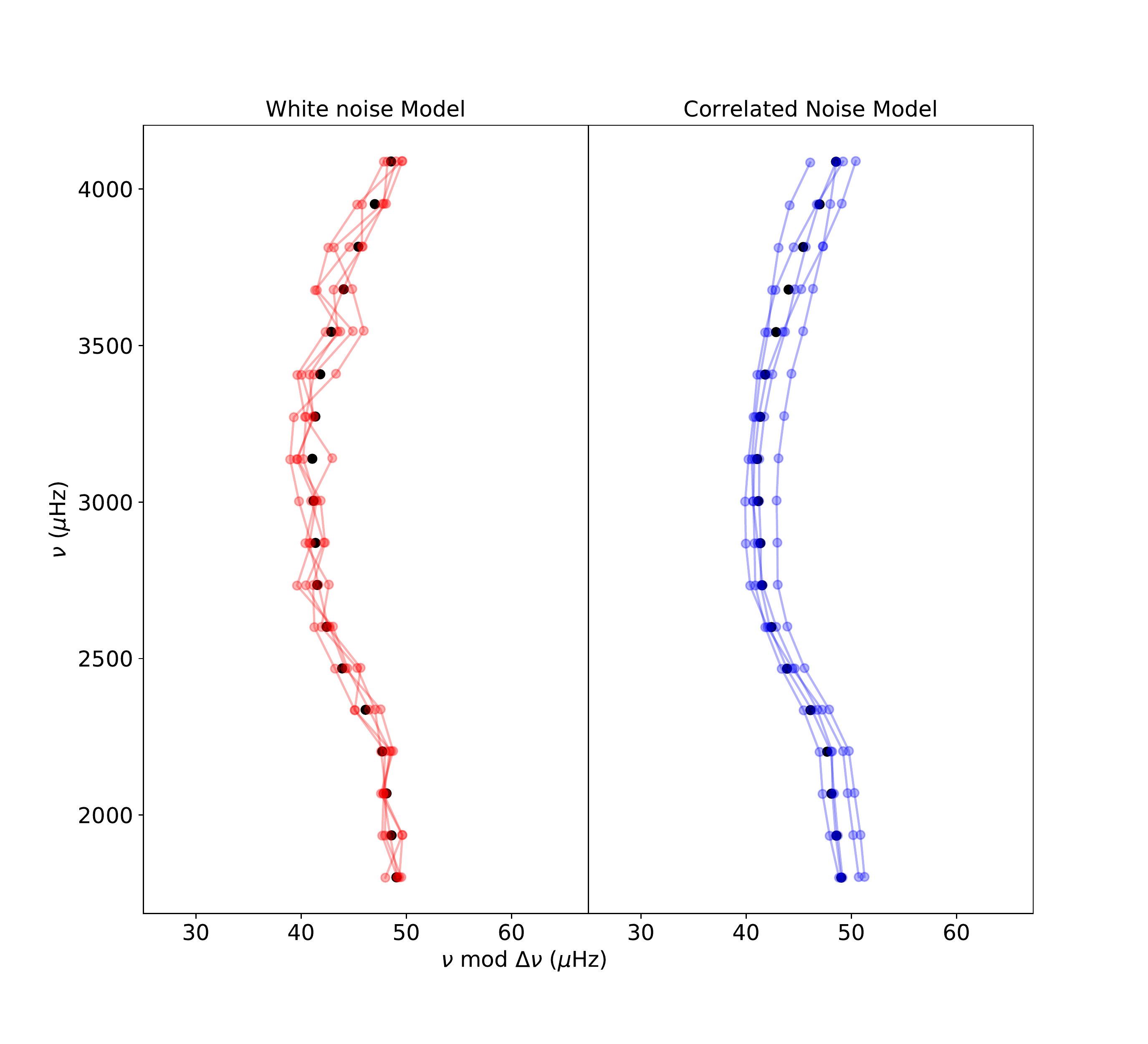}
    \caption{{Mode frequency sets sampled with} the white noise model and the correlated noise model (based on the SE kernel in Eq.~\ref{eq:sek2}) {on the \'Echelle diagram.} Black dots represent a set of mode frequencies of a stellar model computed {with \textsc{MESA} \citep{2011ApJS..192....3P} and \textsc{GYRE} \citep{2013MNRAS.435.3406T} codes. Red dots in the left panel and blue dots in the right panel represent generated noise with the white noise model and the correlated models plus the computed mode frequency.}} 
  \label{fig:model_sys}
\end{figure}

In Figure \ref{fig:model_sys}, we demonstrate some randomly generated noise realizations using Eq~\ref{eq:sek2} based on a set of {computed} model frequencies ($\nu$). {We use a lengthscale of 5$\Delta\nu$ and a variance of 1.0$\mu$Hz. As seen, the predicted frequency sets ($\nu'$) change in a highly correlated way. We also show a set of random samples from the white noise model (WNM) for comparison. The WNM follows a Gaussian distribution and has the same variance value ($\sigma$ = 1.0$\mu$Hz). The predicted frequencies based on the WNM include spiky features, which obviously do not well reflect the true curvature of radial mode frequencies. We suggest here that a CNM based on the SE kernel is the better representation for model systematics because it is more consistent with our prior belief in the nature of the oscillation frequencies.}

\subsection{Determining the Systematic Function with the SE Kernel}

{Now we determine the functional form of CNM. Firstly, we describe the systematic error kernel as}

\begin{equation}\label{eq:sek}
k_{\mathcal{E}} = \sigma_{\mathcal{E}}^2 {\rm exp}\left( - \frac{(\nu + \mu_{\mathcal{E}} - \nu ')^2}{2l_{\mathcal{E}}^2}\right), \\
\end{equation}
{where $\mu_{\mathcal{E}}$ is the mean function, $\sigma_{\mathcal{E}}$ and $l_{\mathcal{E}}$ are the error kernel variance and the error kernel lengthscale, respectively.
Secondly, the systematic uncertainty function contains two terms ($\mathcal{U}_{1}$, and $\mathcal{U}_{2}$) which represent frequency shifts between model grid points. We therefore use two kernels to describe them, written as 
}

\begin{equation}\label{eq:kernel_u}
\begin{split}
&k_{\mathcal{U},1} + k_{\mathcal{U},2}  = \\
& \sigma_{\mathcal{U},1}^2 {\rm exp}\left( - \frac{(\nu - \nu ')^2}{2l_{\mathcal{U},1}^2}\right) + \sigma_{\mathcal{U},2}^2 {\rm exp}\left( - \frac{(\nu - \nu ')^2}{2l_{\mathcal{U},2}^2}\right).\\
\end{split}
\end{equation} 
Here we refer to two kernels  ($k_{\mathcal{U}, 1}$, and $k_{\mathcal{U}, 2}$) as the primary uncertainty kernel and the secondary uncertainty kernel, in which $\sigma_{\mathcal{U}}$ and $l_{\mathcal{U}}$ are the uncertainty kernel variance and the uncertainty kernel lengthscale, respectively.


Finally, we write the full expression for our CNM as 
\begin{equation}\label{eq:s2}
\mathcal{S} = k_{\mathcal{E}} + k_{\mathcal{U}, 1}  + k_{\mathcal{U}, 2} . \\
\end{equation}
The systematic function {considers the systematic errors caused by improper physics in stellar models and the systematic uncertainty due to the model grid steps. To apply the function to stellar mode frequency fits, we need to choose appropriate kernel parameters ($\mu_{\mathcal{E}}$, $l_{\mathcal{E}}$, $\sigma_{\mathcal{E}}$, $l_{\mathcal{U},1}$, $\sigma_{\mathcal{U}, 1}$, $l_{\mathcal{U},2}$, and $\sigma_{\mathcal{U}, 2}$).} In the following section, we will demonstrate how these terms are determined in an actual fitting process.

\subsection{Determining Kernel Parameters}

{We demonstrate the determination of the kernel parameters of the CNM in the case of the Sun-as-a-star. We use the same stellar model grid adopted by \citet{2021MNRAS.505.2427L}. The grid covers parameter ranges (steps) of 0.8 -- 1.2 (0.01) \msolar{} for mass ($M$), -0.5 -- +0.5 (0.1) for metallicity ({\it M/H}), 0.24 -- 0.32 (0.02) for initial helium mass fraction ($Y$), and 1.5 -- 2.5 (0.2) for the mixing-length parameter ($\alpha_{\rm MLT}$).} {Observed frequencies of the Sun are taken from the  \bison\/ network \citep{2017MNRAS.464.4777H}.}

\subsubsection{Free parameters in the Systematic Error Kernel}\label{sec:fp_error_kernel}

Here we discuss the determination of the mean function ($\mu_{\mathcal{E}}$), error kernel lengthscale ($l_{\mathcal{E}}$), and error kernel variance ($\sigma_{\mathcal{E}}$).
The systematic error kernel is normally determined from frequency differences between observations and the best-fitting model.
However, the best-fitting model is unknown to us a priori. Instead, we use specific knowledge of model errors of similar stars to characterise the systematic error and determine three adjusted parameters. In what follows, we justify our choices but note that the method itself could incorporate alternative choices.


Previous studies of the surface correction on stars similar to the Sun provide some useful references for model errors of the Sun.  \citet{2018MNRAS.479.4416C} estimated the surface terms of 66 \kepler{} main-sequence stars. For stars within a parameter range of \teff{} = 5777 $\pm$ 250K and \logg{} = 4.44 $\pm$ 0.5dex, the relative surface corrections at \numax{} ($\delta \nu(\nu_{\rm max})$) vary from -0.0015\numax{} to -0.0045\numax. We use this prior information in the form of a weight $w_{\rm \nu_{max}}$ given by a super Gaussian function with the exponent raised to a power of 10 (flat-top Gaussian function)

 \begin{equation}\label{eq:lm2}
w_{\rm \nu_{max}} = {\rm exp}\left( -\left(\frac{(\delta\nu(\nu_{\rm max}) - \mu_{\nu_{\rm max}})^2}{2\sigma_{\nu_{\rm max}}^2}\right)^{10} \right).
\end{equation}
Here, $\mu_{\nu_{\rm max}}$ and $\sigma_{\nu_{\rm max}}$ are -0.003\numax{} and 0.0015\numax, respectively. This likelihood function returns a weight that is flat when $\delta \nu(\nu_{\rm max})$ is in the $\mu_{\nu_{\rm max}} \pm \sigma_{\nu_{\rm max}}$ range, and quickly drops toward zero at $\mu_{\nu_{\rm max}} \pm 1.5 \sigma_{\nu_{\rm max}}$. 
Moreover, previous studies \citep{2014A&A...568A.123B,2018MNRAS.479.4416C,2020MNRAS.495.3431L} have also inferred that the model frequency errors at the low-frequency range, below $\sim$0.7\numax{}, are not as significant as those in high-frequency range. In sun-like stars, the low-frequency-range errors are on average approximately zero with a small amount of 0.001\numax{} spread. This can be used as another constraint and we describe it as a Gaussian function as

\begin{equation}\label{eq:lm1}
w_{\rm low-\nu} = {\rm exp}\left( -\frac{\widetilde{\delta\nu(\nu_{\rm obs})}^2}{2\sigma_{\rm low-\nu}^2}\right), ({\rm for \;} \nu_{\rm obs} \leq 0.7 \nu_{\rm max}),  
\end{equation}
where $\widetilde{\delta\nu(\nu_{\rm obs})}$ represents the median of {frequency offsets} for given observed frequencies, $\sigma_{\rm low-\nu}$ is adjusted and defines for how much the $\widetilde{\delta\nu(\nu)}$ deviating from zero is sensible. According to earlier studies \citep{2014A&A...568A.123B,2018MNRAS.479.4416C}, we adopt $\sigma_{\rm low-\nu}$ = 0.001\numax{}. 
%

\begin{figure*}
	\includegraphics[width=1.0\columnwidth]{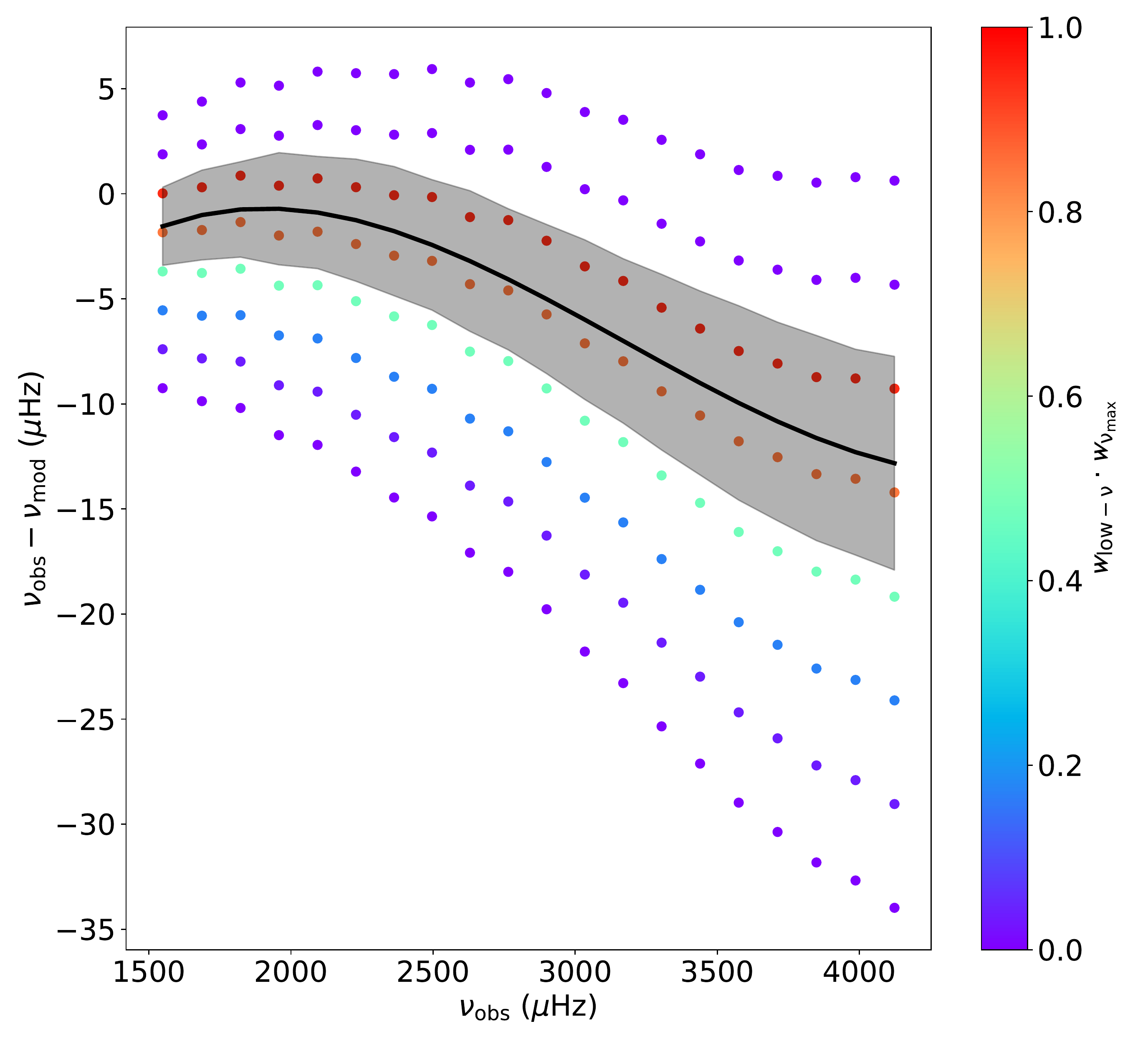}
	\includegraphics[width=1.0\columnwidth]{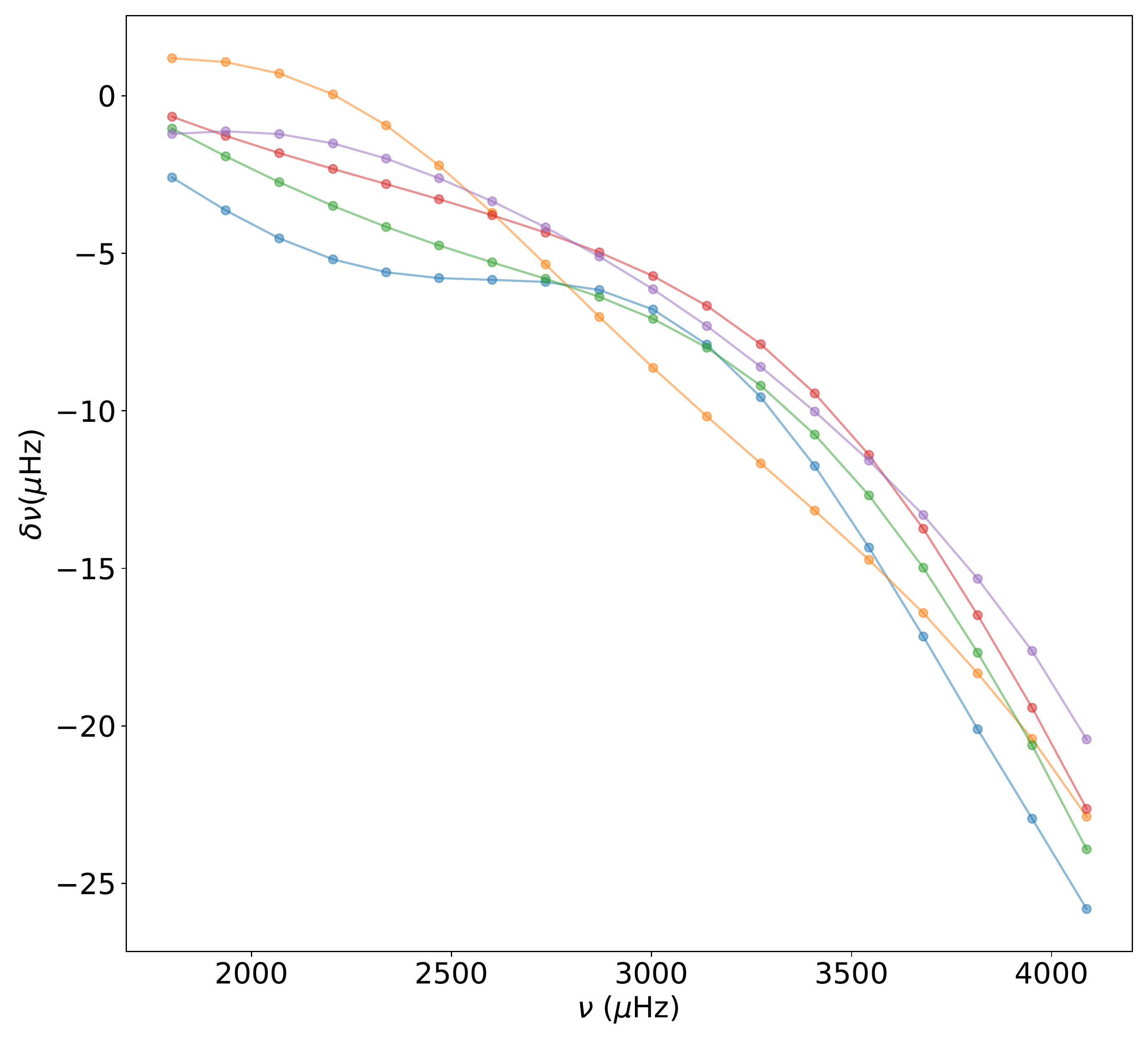}
    \caption{{Left}: {The determination of mean function ($\mu_{\mathcal{E}}$) and variation ($\sigma_{\mathcal{E}}$) of the systematic error kernel ($\mathcal{E}$) while fitting to the Sun-as-a-star.} Dots are frequency differences between observed solar data (from \bison) and theoretical models. These models are from the same evolutionary track with $M$ = 1.00 \smass{}, \yinit{} = 0.26, \feh = 0.0, and \mlt{} = 2.1. The colour code indicates the joint weight determined with the two likelihood functions (Eq. \ref{eq:lm2} and \ref{eq:lm1}).  Note that we use all models on the track to determine the mean function and variation but only plot eight models with a joint weight larger than 0.01. The solid line represents the polynomial function that fits the frequency differences (weighted by the joint weight), and we use this as the mean function. The grey shade indicates the weighted standard deviation which is adopted as the variance. {Right}: Five random draws from the systematic error kernel. The mean function and variance are determined with the method demonstrated in the left panel, and the error kernel lengthscale is 5\Dnu{}.}
  \label{fig:mean_function}
\end{figure*}

Now we demonstrate how these two functions are used to characterise the model errors of the Sun. 
Firstly, we compare observed and model frequencies to calculate frequency differences. 
We then use Eq.~\ref{eq:lm2} and \ref{eq:lm1} to obtain a joint weight ($w_{\rm low-\nu}$$\cdot$$w_{\rm \nu_{max}}$) for each model. 
As a simple example, we use stellar models on an 1\smass{} evolutionary track to demonstrate this estimation in the left panel of Figure \ref{fig:mean_function}. When the joint weight of each model is obtained, we find a couple of potential good-fitting models (with weight value larger than 0.01) on this track and plot their frequency differences as a function of the frequency. The weight distribution indicates the model error is about -5$\mu$Hz at 3,000 $\mu$Hz and goes down to about -10$\mu$Hz at 4,000$\mu$Hz.
Now we could fit a {cubic} polynomial function to all frequency differences in Figure~\ref{fig:mean_function} against the frequency weighted by the joint weight to estimate the mean of frequency offsets (the solid line). Moreover, we calculate the weighted standard deviations of frequency differences (as illustrated by the grey shade). The fitted polynomial function represents the mean estimate of systematic error in these theoretical models, and the standard deviation infer the range for it to vary. We thus use them as the mean function and noise variance in the systematic error kernel. 
%
%
{In our CNM, the lenghtscale determines the degree of correlation between mode frequency noise. By inspecting the surface correction results obtained on the \kepler{} {\it LEGACY} sample \citep{2018MNRAS.479.4416C}, we notice that the frequency offsets between model and observations mostly follow slowly-varying smooth functions (higthly-correlated). This is to say the error kernel lengthscale ought to be much larger than the large separation. 
We test a selection of lengthscale values {and compare the kernel predictions with the surface correction results. We find a suitable lengthscale of 5\Dnu{} because the kernel with this lengthscale value well reproduces both the very smooth surface terms expected for stars with approximately one solar mass and the slightly curved ones, expected for the more massive F-type stars\citep[see Fig. 8 in ][]{2018MNRAS.479.4416C}. 
}
Using the mean function, variance, and lengthscale obtained above, we illustrate some random draws from the systematic error kernel in the right panel of Figure \ref{fig:mean_function}. 

\subsubsection{Free Parameters in the Uncertainty Kernels}\label{sec:fp_uncertainty_kernel}

Now we determine the lengthscales and variances of the two uncertainty kernels. The two lengthscales and variances are different because
the primary kernel describes the general frequency change (a very smooth function of frequency) and the secondary kernel corresponds to the helium glitch signatures (a fast-varying function of frequency). The uncertainty kernels represent the systematic uncertainty at a grid point and their free parameters depend on the local frequency changes between the point and its neighbouring points. Thus, the free parameters in the uncertainty kernels vary for different grid points. 

{Here we use a simple method to} determine free parameters for the grid point at $M$ = 1\smass, [Fe/H] = 0.0, $Y_{\rm init}$ = 0.26, $\alpha_{\rm MLT}$ = 2.1, and $\rho$ = $\rho_{\odot}$. A model grid always contains multiple input dimensions. For the grid in this work, the independent inputs are mass, initial metallicity, initial helium abundance, and initial mixing-length parameters. We therefore inspect the frequency changes for each fundamental input (see details in Figure \ref{fig:app1}).
We find that the primary uncertainty kernel needs a large lengthscale to make it act as a very smooth function. {We test different lengthscale from 10\Dnu{} to 30\Dnu{} and adopt $l_{\mathcal{U}, 1}$ =20\Dnu{} because it best recovers the general shape of frequency changes in Figure \ref{fig:app1}}. On the other hand, we use a small lengthscale for the secondary uncertainty kernel to match the quick variation of the glitch signature. 
{By inspecting the second-order variations in Figure \ref{fig:app1}, we notice that the lengths of the `wiggles' are mostly between 2--3 radial orders. The choice of lengthscale needs to be in a similar range to reproduce the fast variations. We hence test different lengthscale values from 1\Dnu{} to 4\Dnu{}} and find the kernel with $l_{\mathcal{U}, 2}$ = 2\Dnu{} best recovers the test case. 
Considering the frequency resolution for all input dimensions, we find an average varying range is $\sim$0.75$\mu$Hz at 0.5\numax, $\sim$1.5$\mu$Hz at \numax, and $\sim$2.0$\mu$Hz at 1.5\numax. The variance is hence frequency dependent and we use $\sigma_{\mathcal{U}, 1}$ = $1.5\nu_{\rm obs}/\nu_{\rm max}$. 
The secondary kernel variance is also frequency-dependent given that the signature of the helium glitch is a damped sine wave. As shown in Figure~\ref{fig:model_sys_glitch}, the secondary uncertainty is up to $\sim$0.4$\mu$Hz at 0.5\numax{} and gradually reduces to $\sim$ 0 at 1.5\numax. We therefore set up the secondary variance as $\sigma_{\mathcal{U}, 2}$ = 0.1($\nu_{\rm max}/\nu_{\rm obs}$)$^{2}$.
{Note that the method of measuring frequency differences and estimating the kernel variation is fairly rough. Firstly, the kernel variation is not uniform through a model grid. Secondly, we measure frequency changes in each input dimension independently, but the four fundamental inputs highly degenerate. The grid resolution of oscillation frequency and the free parameters of the uncertainty kernels should be solved as multiple-dimesions problems. We leave these in future studies. }

%

In Figure \ref{fig:kernel_u}, we use the two uncertainty kernels to generate some frequency noise as a function of frequency. As seen, the primary uncertainty kernel generates very smooth and highly correlated noise, and what the secondary uncertainty kernel provides is similar to the damped sine wave. The combination of the two kernels gives reasonable predictions of frequency noise between stellar model grids. 

\begin{figure}
	\includegraphics[width=1.0\columnwidth]{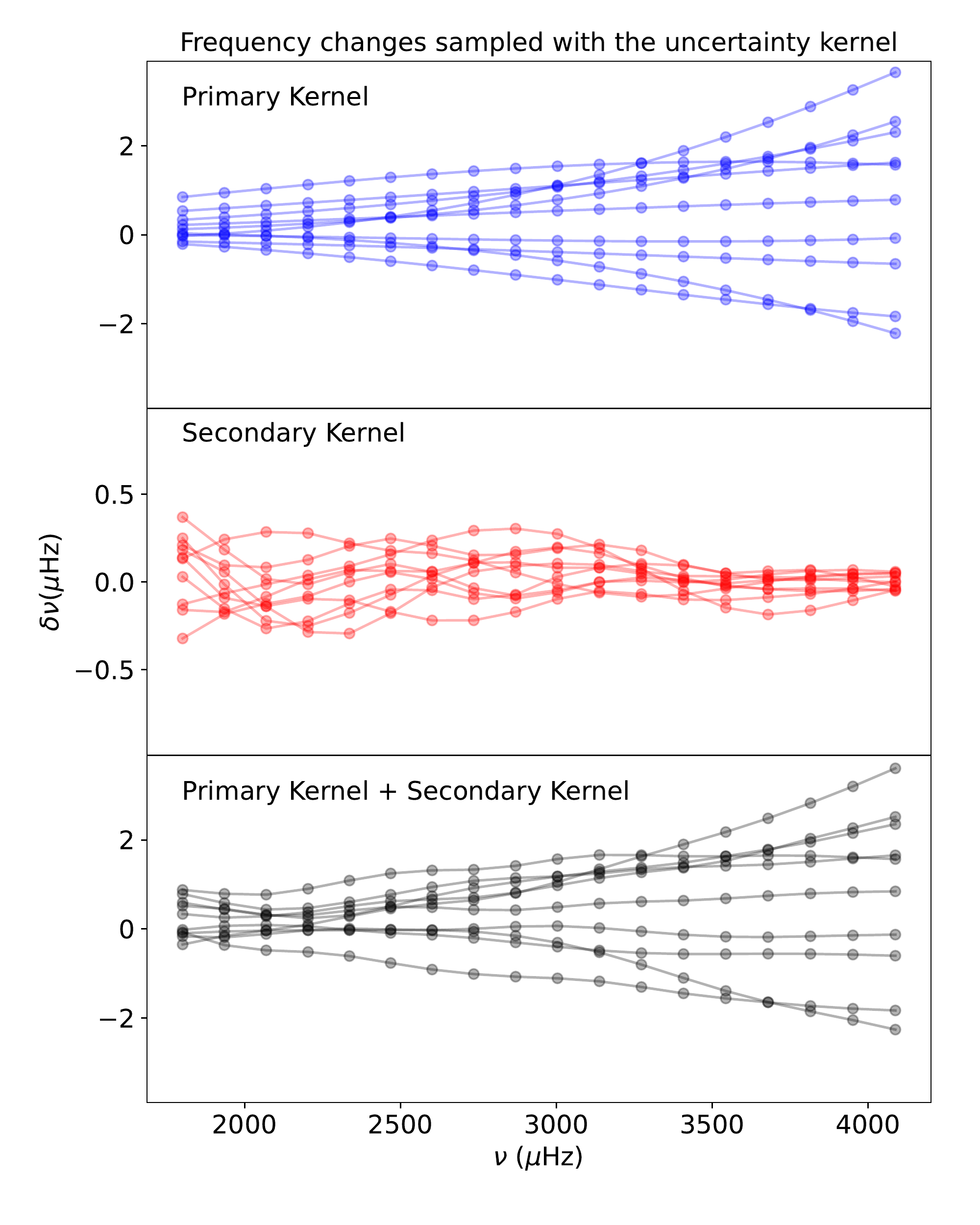}
    \caption{{Generated frequency noise using the primary uncertainty kernel (top), the secondary uncertainty kernel (middle), and the combination of two kernels (bottom). The primary kernel has a lengthscale of 20\Dnu{} and a frequency-dependent variance of 1.5$\nu_{\rm obs} / \nu_{\rm max}$. The secondary kernel, which describes the change in glitch signature, has a lengthscale of 2\Dnu{} and a variance of 0.1($\nu_{\rm max} / \nu_{\rm obs}$)$^{2}$. In each panel, we demonstrate 10 randomly generated sets of frequency noise, i.e., the ($\nu$ - $\nu '$) in the systematic uncertainty kernel. Note that we use observed solar frequencies as $\nu$ when generating the noise in this plot.}
    } 
  \label{fig:kernel_u}
\end{figure}

\subsection{Likelihood Function}

In this section, we discuss the likelihood function that works for the CNM fitting. We start with the likelihood function used in traditional methods. When no systematics are considered and errors on the observed frequencies are assumed to be uncorrelated, it is normally written as 

\begin{equation}\label{eq:lk1}
    \mathcal{L}_{\rm} = {\rm exp}\left( - \frac{(\nu_{\rm obs} - \nu_{\rm mod}  + \delta\nu_{\rm sc})^2}{2\sigma^2_{\nu_{\rm obs}} } \right), \\
\end{equation}
where the subscripts `obs' and `mod' stand for the observed and the model quantities. The term $\delta\nu_{\rm sc}$ represents the surface correction to model frequencies. 
When we consider model systematic uncertainties and treat them as white noises, the likelihood function becomes 

\begin{equation}\label{eq:lk2}
    \mathcal{L}_{\rm WNM} = {\rm exp}\left( - \frac{(\nu_{\rm obs} - \nu_{\rm mod}  + \delta\nu_{\rm sc})^2}{2(\sigma^2_{\nu_{\rm obs}}  +  \sigma^2_{\nu_{\rm sys}} ) } \right), \\
\end{equation}
where $\sigma_{\nu_{\rm sys}}$ represents the white systematic noise. 

In Figures \ref{fig:mean_function} and \ref{fig:kernel_u}, we generate $\nu'$ for justifying our choices of kernel parameters. In the fitting process, generating $\nu'$ with some samplers (e.g., MCMC) and fitting $\nu'$ to observations is applicable but computationally expensive. In the fitting procedure, we could simply use likelihood functions instead of sampling $\nu'$. Because a GP kernel is essentially a covariance matrix for a multivariate Normal distribution, whose probability function follows a multivariate normal distribution. Thus, we can describe CNM as a multivariate normal distribution in the fits and write the likelihood function as

\small
\begin{equation}\label{eq:lk3}
\begin{split}
   & \mathcal{L}_{\rm CNM, N} \propto 
   &  {\rm exp}\left(-\frac{1}{2}(\nu_{\rm mod} + \mu_{\mathcal{E}} - \nu_{\rm obs})^T \sum{} ^{-1}(\nu_{\rm mod} + \mu_{\mathcal{E}} - \nu_{\rm obs}) \right). \\
\end{split}
\end{equation}
\normalsize
The subscript `N' stands for the Normal distribution, and the $\sum{}$ represents the covariance matrix {based on the SE kernel} 

\begin{equation}\label{eq:lk3_sub}
\sum =k_{\rm com}(\nu_{\rm mod}, \nu_{\rm mod}) + I \cdot  \sigma^2_{\nu_{\rm obs}}.
\end{equation}
{ The term $k_{\rm com}(\nu_{\rm mod}, \nu_{\rm mod})$ is the combination of three kernels which describe correlated noise.  The term $I \cdot  \sigma^2_{\nu_{\rm obs}}$ ($I$ is the identity matrix) describes the observed uncertainty of mode frequencies which should be white noise. It should be noted that the $k_{\rm com}(\nu_{\rm mod}, \nu_{\rm mod})$ term does not contain $\mu_{\mathcal{E}}$ in the systematic error kernel as we already consider $\mu_{\mathcal{E}}$ when comparing model and observed frequencies as shown in Eq. \ref{eq:lk3}.}

%
%
%
Moreover, we could consider in addition some potential errors which are unknown or unpredictable. In some cases, a few per cent of observed mode frequencies can be misreported or poorly measured. {Some noisy spikes in the oscillation power spectrum could be misclassified as modes; mixed modes close to the $\ell = 0$ ridge could be misidentified as radial modes; and frequency uncertainties could be underestimated for un-resolved modes.} Due to these additional errors, the probability distribution should contain a small fraction of `exceptions'. 
{The $t$-distribution is a good option for dealing with these cases.} The probability density function of the $t$-distribution is similar to the normal distribution but has long and fat tails at both sides. The probability in the tail region represents the chance of a mode frequency being misreported. 
The likelihood function following the multivariate t-distribution is written as 

\small
\begin{equation}\label{eq:lk4}
\begin{split}
   & \mathcal{L}_{\rm CNM, {\it t}} \propto \\
    & \left(1 + \frac{1}{d}(\nu_{\rm mod} - \nu_{\rm obs})^T \sum{} ^{-1}(\nu_{\rm mod} - \nu_{\rm obs}) \right)^{-(d+p)/2}. \\
\end{split}
\end{equation}
\normalsize
The subscript `$t$' represents the $t$-distribution, $p$ is the dimension of the vector of frequencies, and $d$ is the number of degrees of freedom that determines the possibility of incorrect measurement. The degrees of freedom $d = 2$ corresponds to a $\sim$10\% probability in the tail regions (outside 3 times half width at half maximum) and we adopt this value in the following analysis. 



\section{Application in asteroseismic fitting}\label{sec:results}

\subsection{Fitting a fake model star}

To test whether this new fitting method better recovers the truths of stellar parameters. We fit to a fake model star which has similar surface properties to the Sun.
The fake star is computed with same input physics but with off-grid input fundamental parameters, which are M = 1.005\msolar, \yinit{} = 0.256, \mlt{} = 1.99. The true age, radius, and mean density are $\tau$ = 3.99Gyr, $R$ = 1.007\rsolar, and $\overline{\rho}$ = 0.984$\overline{\rho}_{\odot}$, respectively.
Classical `observed' quantities are \teff{} =  5771K, \logg{} = 4.43dex, \feh{} = -0.05dex. We adopt observed uncertainties of $\pm$50K for \teff{}, $\pm$0.1dex for \logg{}, and  $\pm$0.1dex for \feh. Radial mode frequencies for $n$ = 12 -- 30 are selected as seismic constraints. The global seismic parameters are \numax{} = 3058$\mu$Hz (calculated with the scaling relation given by \citet{1995A&A...293...87K}) and \Dnu{} = 134.4$\mu$Hz (from the linear fitting of radial mode frequencies). A uniform observed uncertainty of  $\sigma_{\nu_{\rm obs}} = 0.5\mu$Hz for mode frequencies is applied.
{Regarding the free parameters of the kernels, we use fixed parameters determined in Sections~\ref{sec:fp_uncertainty_kernel} and \ref{sec:fp_error_kernel} in all following tests.}

Because the star is a fake model star, no model error is included. The systematic function only contains the uncertainty term. 
We consider three cases for comparison. We do not consider the systematics for the first case and use Eq. \ref{eq:lk1}  as the likelihood function. The second case includes the white systematic noise and {Eq. \ref{eq:lk2} as the likelihood function}. In the third case, we apply the correlated noise model and use the multivariate normal distribution (Eq. \ref{eq:lk3}) to determine the likelihood function. Note that we use the same Maximum Likelihood Estimation (MLE) to fit classical observed frequencies in all these three cases. The posterior distribution is the joint probability of likelihoods from the classical and the seismic fits.


We start with the first and the second cases which either ignore model systematics or treat them as white noise.
Frequency-dependent systematic uncertainties as $\sigma_{\rm sys}$ = $1.5 \nu_{\rm obs}/\nu_{\rm max}$ are used in the second case.
 Figure~\ref{fig:corners} shows posterior distributions of five fundamental parameters. Posterior distributions for the first case are spiky because the model grid is significantly under-sampled, leading to poor accuracy and precision. When the white systematic uncertainty is considered, we obtain continuous posterior distributions and more reliable estimates of stellar parameters. 
{We then apply the CNM as described in Section~\ref{sec:method} in the fits and illustrate the fitting results at the bottom in Figure~\ref{fig:corners}. }Comparing with above traditional methods, we find significant improvements in estimated precisions for mass, radius, and age because CNM is a more principled statistical treatment for the systematics. 

\begin{figure*}
\includegraphics[width=1.\columnwidth]{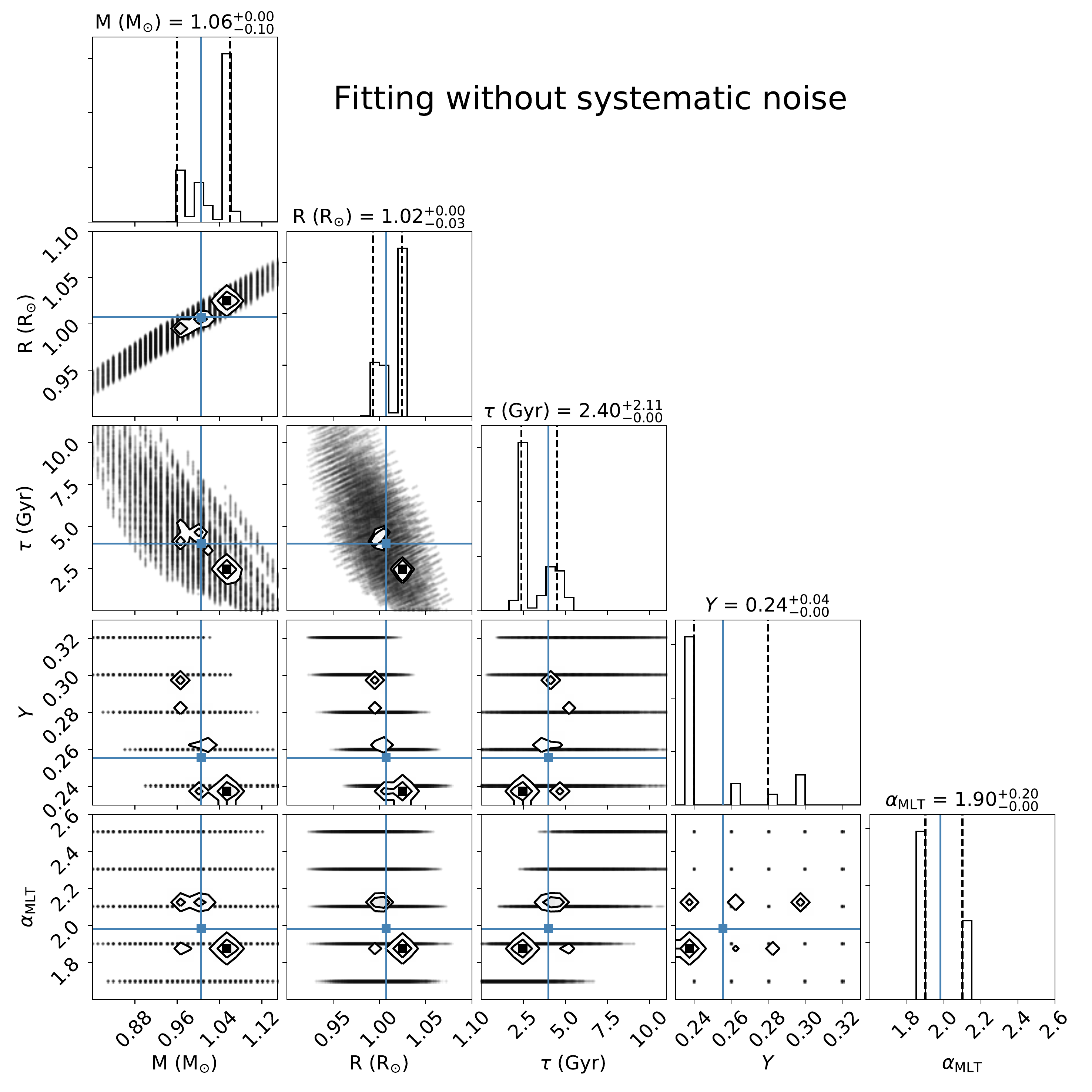}
\includegraphics[width=1.\columnwidth]{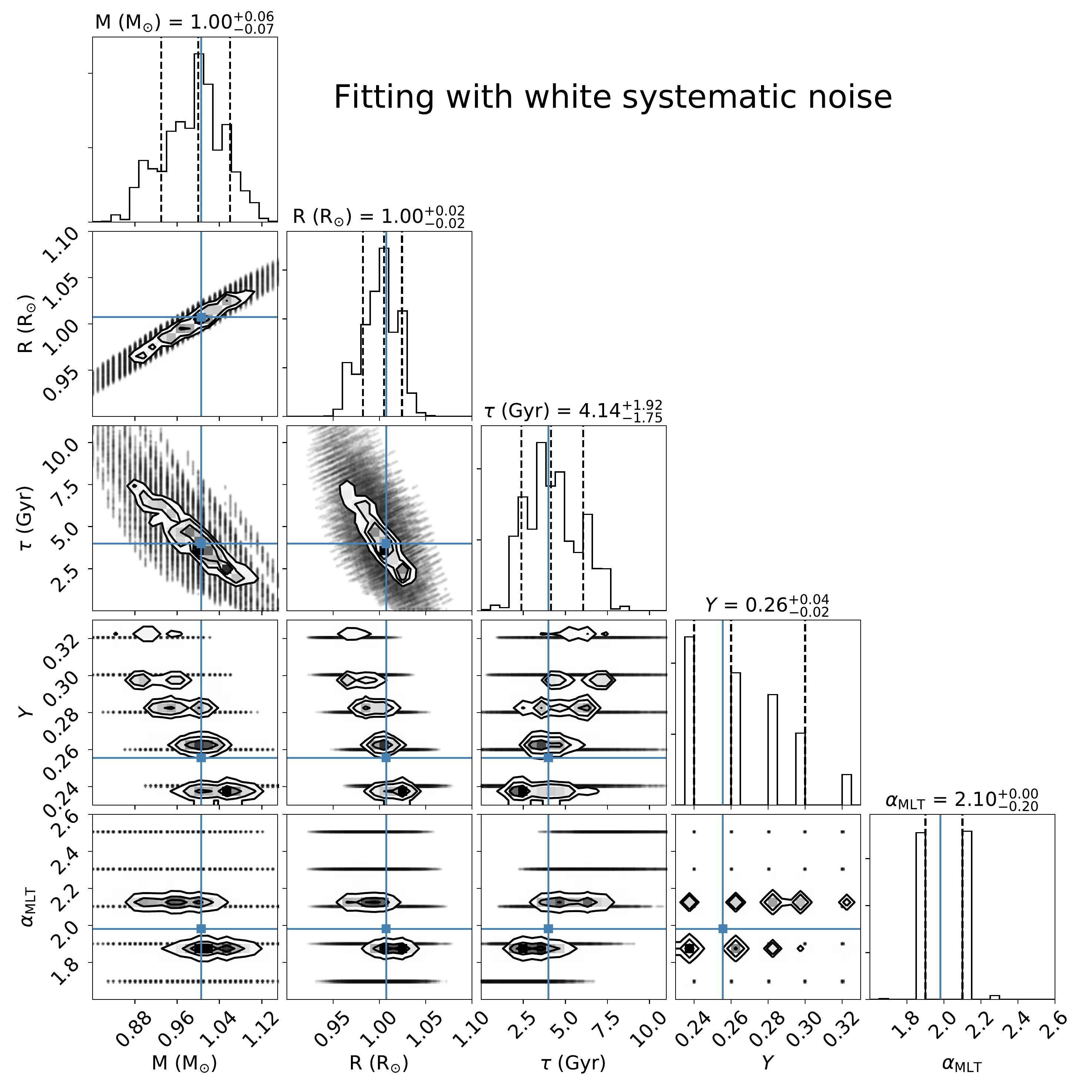}
\includegraphics[width=1.\columnwidth]{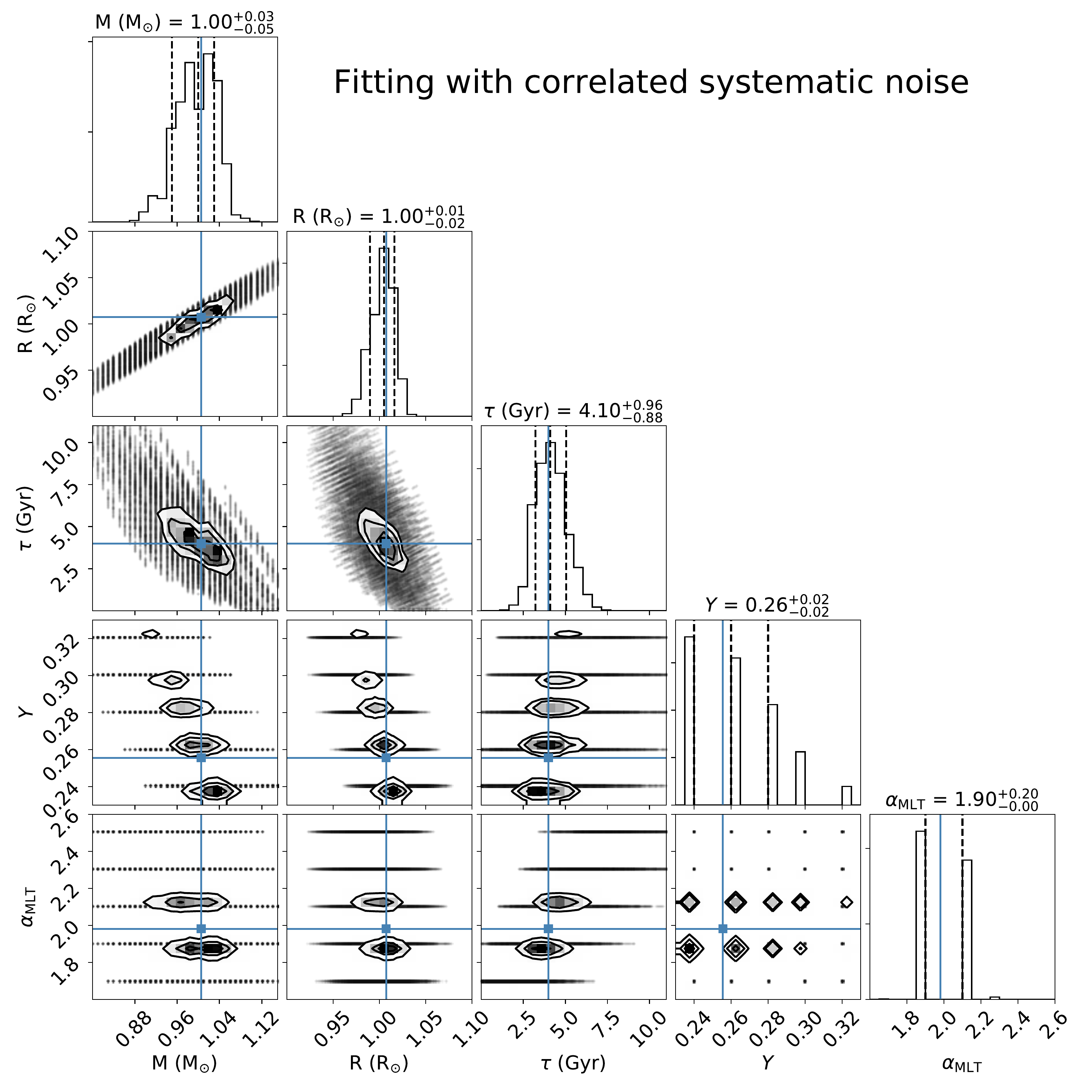}
    \caption{Probability distributions of five star parameters of the fake star on the \textsc{Corner} \citep{corner} plot. Results include three cases which are fits without any model systematics (top left), with white systematic noise (top right), with a correlated noise model (bottom). {The input values of the five stellar parameters are} indicated by blue solid lines.} 
  \label{fig:corners}
\end{figure*}

\subsection{Fitting a realistic fake star}

As a further test of our method, we make the above fake star more realistic by adding the surface term and some random observed noise to oscillation frequencies. To be specific, we refer to it as the `realistic fake star'. 
We add the solar surface term using the correction formula given by \citet{2008ApJ...683L.175K} with the two adjusted parameters as $a$ = -4.73 and $b$ = 4.90. Some random noise following the Gaussian distribution ($\sigma_{\rm obs}$ = 0.5$\mu$Hz) is added to the frequencies. We generate several sets of `observed' frequencies and choose the one with a few frequencies obviously deviating from the asymptotic relation (as shown on the top left in Figure~\ref{fig:fake_real_star}). This is to mimic the case of poorly measured modes. All the observed quantities are the same except \Dnu{} because adding the surface term changes its value. We fit the new mode frequencies with a linear function and obtain an observed \Dnu{} of 133.4$\mu$Hz. 

The full systematic function is applied in the fitting. We first fit to three classical observed quantities (\teff{}, \logg{} and \feh) using the MLE method and select models with likelihood greater than $10^{-4}$.
We then determine the mean function ($\mu$) and the variance ($\sigma_{\mathcal{E}}$) for the error kernel ($\mathcal{K}_{\mathcal{E}}$) with the selected models. As illustrated on the top right in Figure~\ref{fig:fake_real_star}, we calculate the prior likelihoods with Eqs~\ref{eq:lm2} and \ref{eq:lm1} and calculate the weighted median and the weighted standard deviation for observed frequencies. 
The uncertainty kernel is defined in the same way as in the fits to the fake model star. 

We fit models to the data and calculate the likelihood with the multivariate Normal distribution (Eq~\ref{eq:lk3}) as well as the multivariate $t$-distribution (Eq.~\ref{eq:lk4}). We show posterior distributions for the two cases at the bottom of Figure~\ref{fig:fake_real_star}. {Estimated mass, radius, and age for both cases are consistent with the input parameters. Comparing results with two different likelihood functions}, we find better age precision from the fits with multivariate $t$-distribution but no obvious improvements for estimated mass and radius. From the joint distributions, we find larger age spreads against the helium fraction and the mixing-length parameter for the Normal-distribution case. This is because those poorly-measured modes form some small-scale curvatures which affect the glitch signature. 
This is to say, those bad modes weaken the indications of mode frequencies to the helium abundance. The Normal distribution likelihood function does not consider any exceptions {in oscillation frequencies and hence treat those mode shifts as helium glitches, leading to larger spreading in the posterior of the helium fraction.} On the other hand, the likelihood function with $t$-distribution can properly interpret them as potential errors and hence avoid over-explaining the data.

 \begin{figure*}
\includegraphics[width=1.\columnwidth]{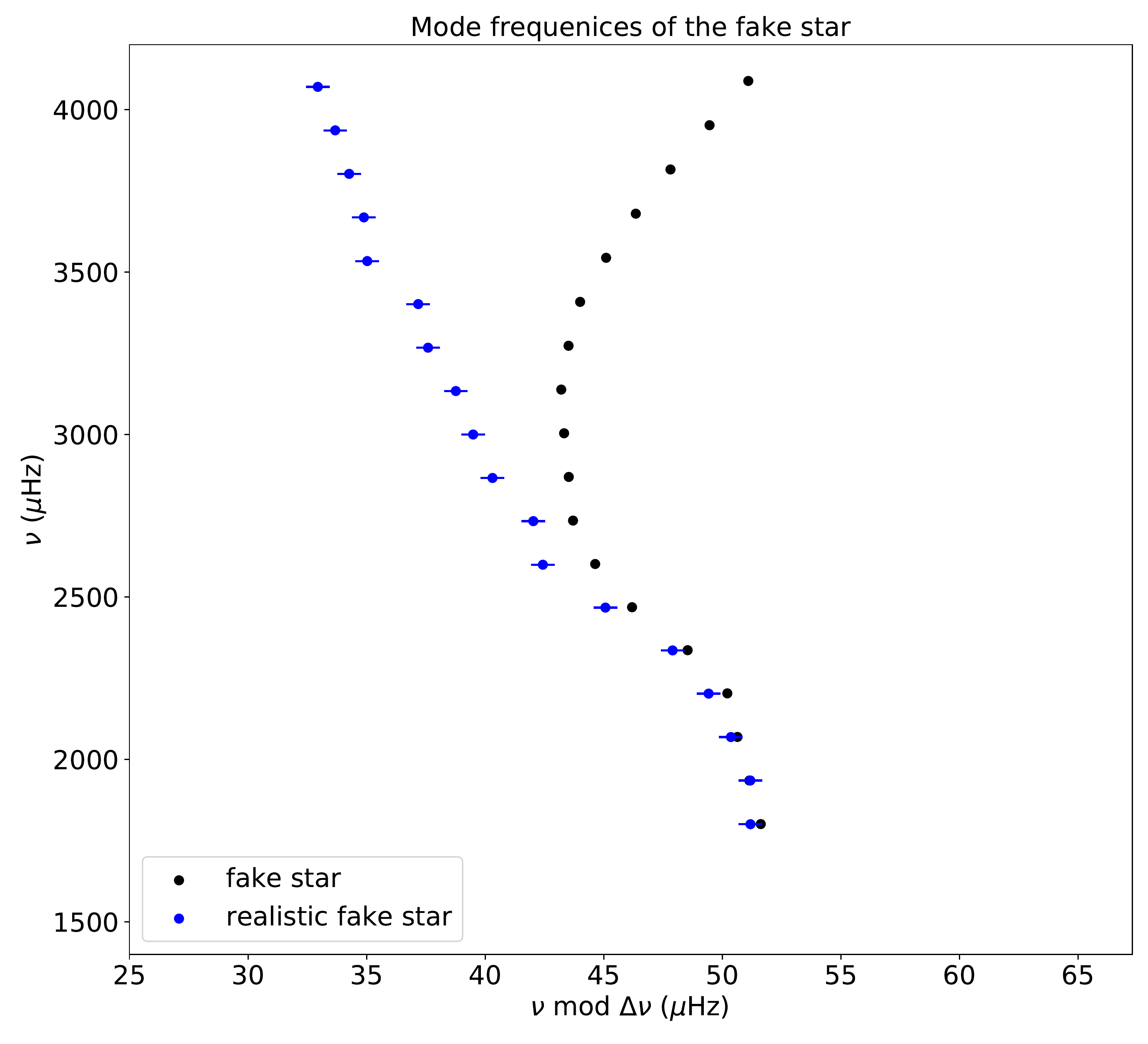}
\includegraphics[width=1.\columnwidth]{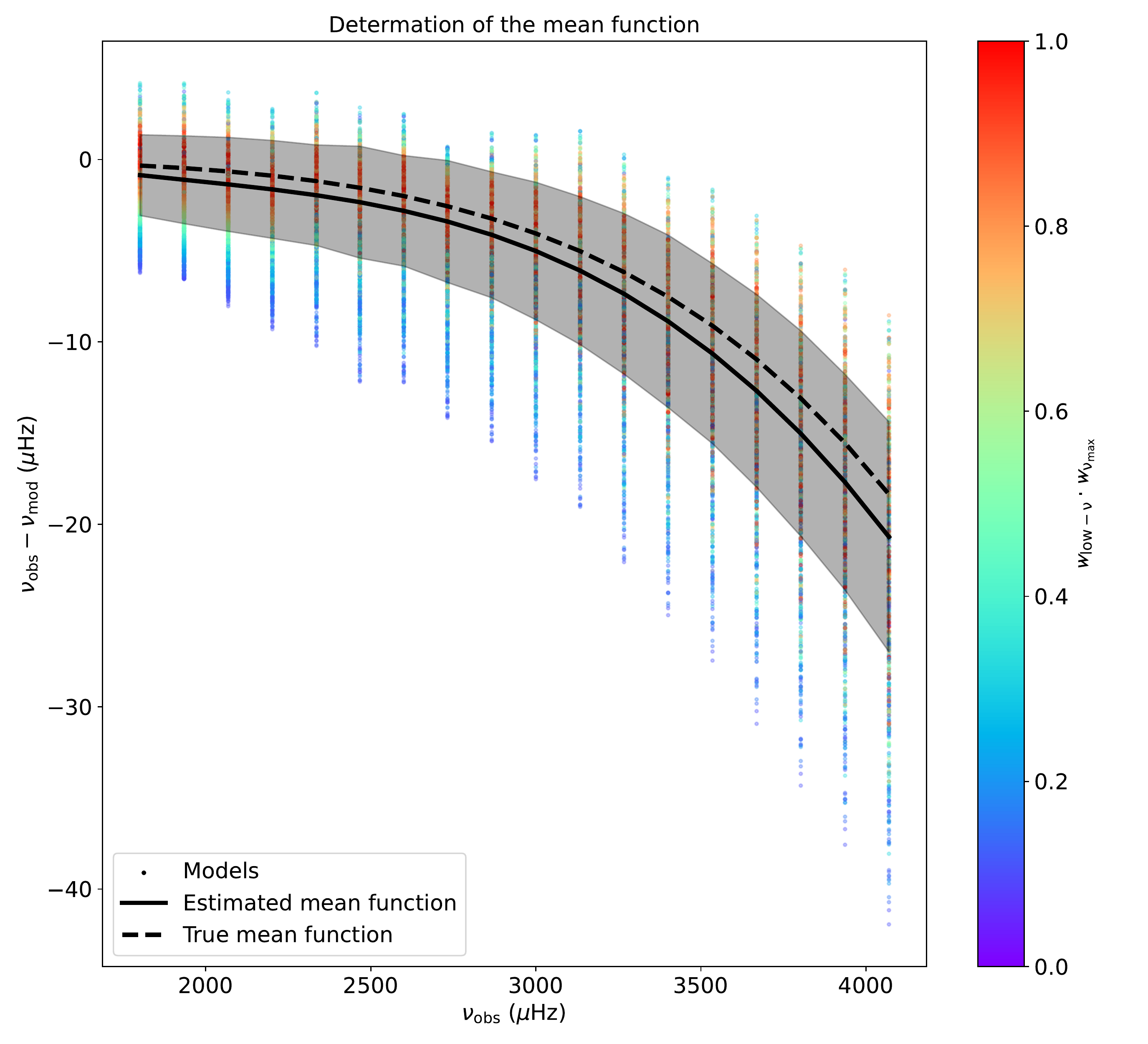}
\includegraphics[width=1.\columnwidth]{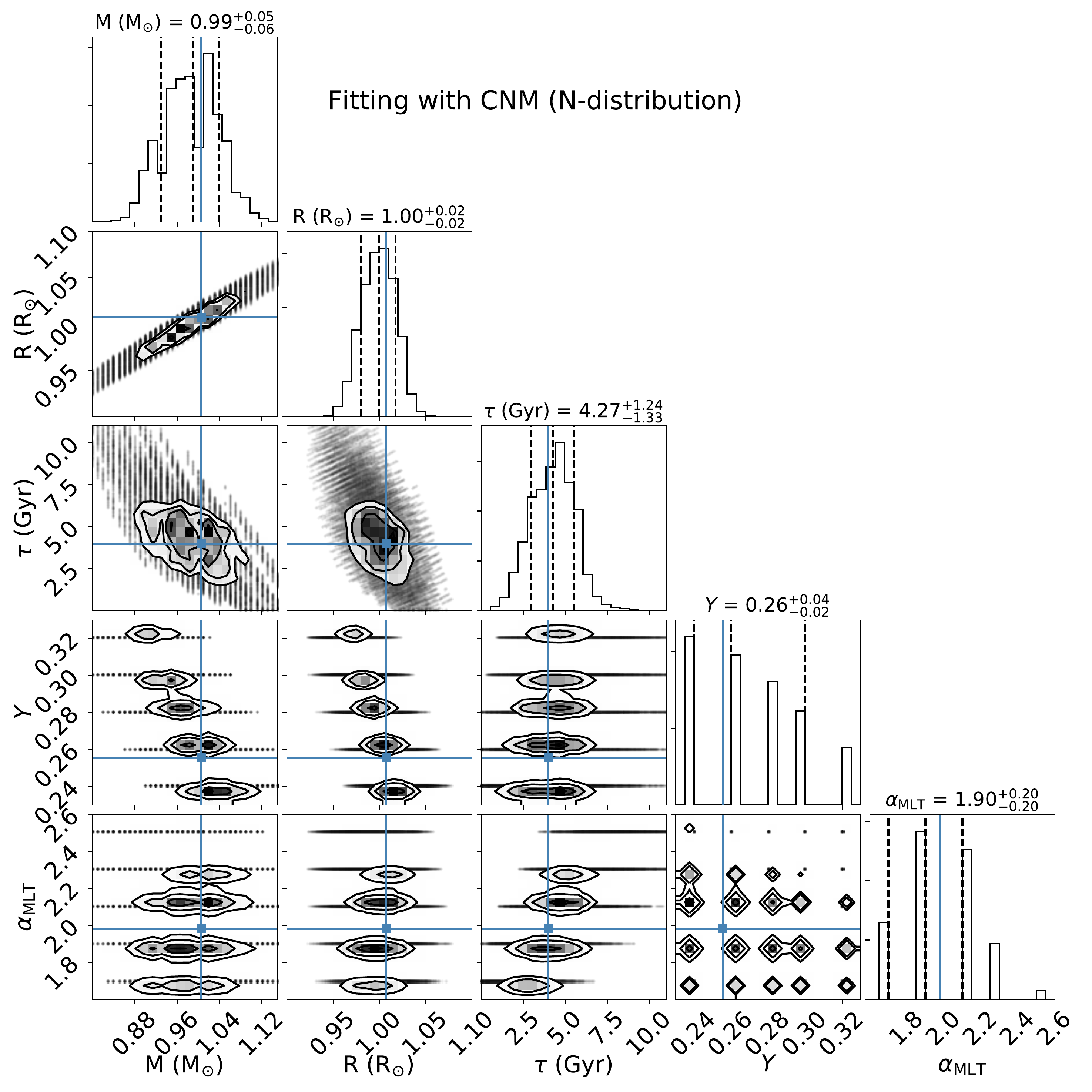}
\includegraphics[width=1.\columnwidth]{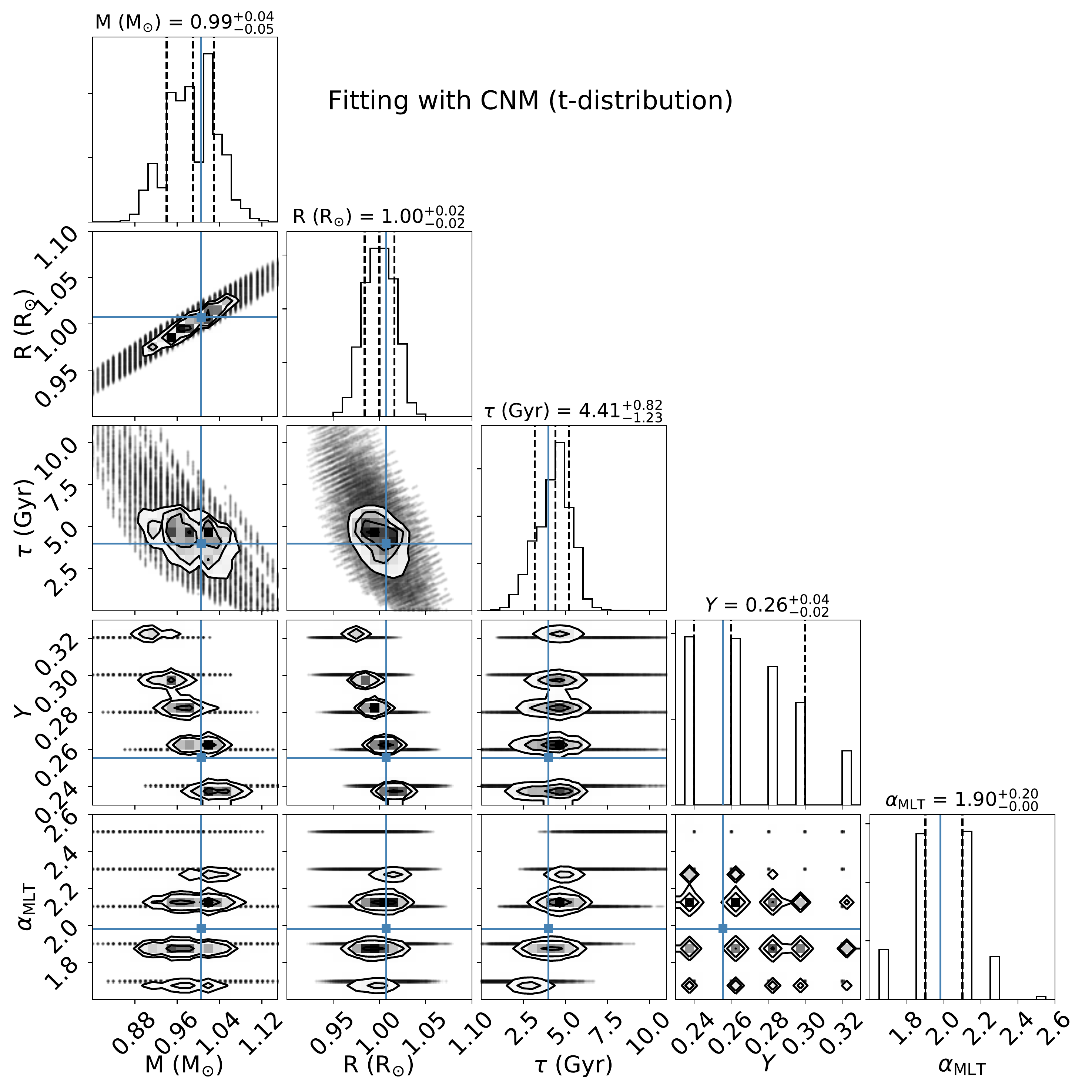}
    \caption{Fitting procedure for the `realistic fake star'. Top left: `observed' radial mode frequencies (blue dots) generated with model frequencies (black dots) plus the solar surface term and some random noises. Top right: the determination of mean function of model errors ($\mu$ in Eq.~\ref{eq:s2}). Dots indicates frequency offsets between models and observations, and colour code represent the joint probability of two priors ($\mathcal{L}_{\rm low-\nu} \cdot \mathcal{L}_{\rm \nu_{max}}$) of each model. Black solid line and grey shade are weighted median and weighted standard deviation of frequency offsets. Black dashed line stands for true model errors. Bottom left: posterior distributions of five stellar parameters based on the multivariate Normal distribution (Eq.~\ref{eq:lk3}) on the \textsc{Corner} plot. Truths are presented by blue lines. Bottom right: same as the left, but for the the multivariate $t$-distribution (Eq.~\ref{eq:lk3}) case. } 
  \label{fig:fake_real_star}
\end{figure*}


%
%

\section{Discussions and Conclusions}\label{sec:conclusion}

We propose a correlated noise model based on a Gaussian Process kernel (covariance function) to better describe model systematics in stellar models constrained by asteroseismic mode frequencies. The work is motivated by poor treatments of model systematics in theoretical mode frequencies {which undermine} modelling solutions. The model systematics include errors caused by improper physics and uncertainties due to the grid resolution. We use a {GP kernel} because it has infinitely many derivatives in its prior to represent all possible frequency variations {between the points of a model grid}. We show that this CNM generates mode frequencies that better match our expectations than the white noise model (Figure~\ref{fig:model_sys}). In practice, we describe the systematics as a mean function of frequency offsets and several kernels. We manage to use these kernels to reproduce frequency variations at different scales by tuning the two free parameters, i.e., kernel lengthscale and kernel variance. We apply the method to fitting a simulated set of model frequencies. In this work, fits with this new CNM outperform the other two traditional methods which either ignores the systematics or treat them as white noises. We also suggest using the $t$-distribution likelihood function in the fitting to cope with potentially misreported mode frequencies. Our testing shows that the $t$-distribution likelihood function better recovers the properties of the simulated star compared with the Normal-distribution.

The new fitting approach includes a better description for the model systematics and hence improves the reliability of modelling solutions. This is a novel alternative to account for the effect of the limited resolution of the grid. It mitigates the issues in seismic modelling, as found by \citet{2021MNRAS.508.5864C}, that when the errors on the frequencies are not inflated the uncertainties on the inferred stellar properties are often {underestimated}. We also treat the surface term (model errors) as a mean function plus a covariance function instead of parameterising a specific correction formula. {The method can be applied to any established model grid.} It is fast and statistically-sound in the grid-based framework and hence suitable for modelling a large sample of stars. The method could be useful for the \plato{} stellar analysis pipeline \citep[][]{2021arXiv211106666G}, which is developed for fast determinations of stellar parameters in the core program of the mission.
This work only demonstrates the application to radial modes, but the method is extendable to all acoustic modes and also possibly to mixed modes. 

There are also some limitations. Although the correlated noise model is a reasonable prediction of model systematics, fitting with it is still not as good as interpolating the grid (assuming the error on interpolation is significantly smaller than the frequency errors). The fitting approach is therefore not the best option when precision is essential. The second limitation is in the determination of the mean function appearing in the systematic error kernel. It relies on previous studies on similar stars to give reliable priors. Here we obtain a good mean function for the example star, because its parameters fit in many well-studied Sun-like stars. To apply this method to other types of stars, a well-studied star sample that covers wide parameter ranges is required for estimating the mean function. Moreover, we note that the functional form of kernels can change for different input physics or free parameters of model grids.} Analysing model systematics is required before applying the method to model grids. {In future work, we will introduce additional machine-learning tools to learn the model systematics across the HR diagram to determine the functional form in a robust way. This will make the method applicable for many stars without customising the GP kernels.

\section*{Acknowledgements}
This paper has received funding from the European Research Council (ERC) under the European Union’s Horizon 2020 research and innovation programme (CartographY GA. 804752). 
This work is supported by the Joint Research Fund in Astronomy (U2031203) under cooperative agreement between the National Natural Science Foundation of China (NSFC) and Chinese Academy of Sciences (CAS), and NSFC grants (12090040, 12090042). 
This work is also supported by the Fundamental Research Funds for the Central Universities.
MBN acknowledges support from the UK Space Agency. MSC acknowledges the support by FCT/MCTES through the research grants UIDB/04434/2020, UIDP/04434/2020 and PTDC/FIS-AST/30389/2017 and the contract CEECIND/02619/2017, and by FEDER - Fundo Europeu de Desenvolvimento Regional through COMPETE2020 - Programa Operacional Competitividade e Internacionalização (grant: POCI-01-0145-FEDER-030389).



\section*{Data availability}

The data underlying this article are available in GitHub, at \url{https://github.com/litanda/csnm_fitting}. 

\bibliographystyle{mnras}
\bibliography{ref} 

\begin{thebibliography}{}
\makeatletter
\relax
\def\mn@urlcharsother{\let\do\@makeother \do\$\do\&\do\#\do\^\do\_\do\%\do\~}
\def\mn@doi{\begingroup\mn@urlcharsother \@ifnextchar [ {\mn@doi@}
  {\mn@doi@[]}}
\def\mn@doi@[#1]#2{\def\@tempa{#1}\ifx\@tempa\@empty \href
  {http://dx.doi.org/#2} {doi:#2}\else \href {http://dx.doi.org/#2} {#1}\fi
  \endgroup}
\def\mn@eprint#1#2{\mn@eprint@#1:#2::\@nil}
\def\mn@eprint@arXiv#1{\href {http://arxiv.org/abs/#1} {{\tt arXiv:#1}}}
\def\mn@eprint@dblp#1{\href {http://dblp.uni-trier.de/rec/bibtex/#1.xml}
  {dblp:#1}}
\def\mn@eprint@#1:#2:#3:#4\@nil{\def\@tempa {#1}\def\@tempb {#2}\def\@tempc
  {#3}\ifx \@tempc \@empty \let \@tempc \@tempb \let \@tempb \@tempa \fi \ifx
  \@tempb \@empty \def\@tempb {arXiv}\fi \@ifundefined
  {mn@eprint@\@tempb}{\@tempb:\@tempc}{\expandafter \expandafter \csname
  mn@eprint@\@tempb\endcsname \expandafter{\@tempc}}}

\bibitem[\protect\citeauthoryear{{Aguirre B{\o}rsen-Koch} et~al.,}{{Aguirre
  B{\o}rsen-Koch} et~al.}{2022}]{2022MNRAS.509.4344A}
{Aguirre B{\o}rsen-Koch} V.,  et~al., 2022, \mn@doi [\mnras]
  {10.1093/mnras/stab2911}, \href
  {https://ui.adsabs.harvard.edu/abs/2022MNRAS.509.4344A} {509, 4344}

\bibitem[\protect\citeauthoryear{{Appourchaux} et~al.,}{{Appourchaux}
  et~al.}{2012}]{2012A&A...543A..54A}
{Appourchaux} T.,  et~al., 2012, \mn@doi [\aap] {10.1051/0004-6361/201218948},
  \href {https://ui.adsabs.harvard.edu/abs/2012A&A...543A..54A} {543, A54}

\bibitem[\protect\citeauthoryear{Ball}{Ball}{2017}]{ball2017surface}
Ball W.~H.,  2017, in EPJ Web of Conferences. p. 02001

\bibitem[\protect\citeauthoryear{{Ball} \& {Gizon}}{{Ball} \&
  {Gizon}}{2014}]{2014A&A...568A.123B}
{Ball} W.~H.,  {Gizon} L.,  2014, \mn@doi [\aap] {10.1051/0004-6361/201424325},
  \href {https://ui.adsabs.harvard.edu/#abs/2014A&A...568A.123B} {568, A123}

\bibitem[\protect\citeauthoryear{{Borucki} et~al.,}{{Borucki}
  et~al.}{2009}]{2009IAUS..253..289B}
{Borucki} W.,  et~al., 2009, in {Pont} F.,  {Sasselov} D.,   {Holman} M.~J.,
  eds,  Vol. 253, Transiting Planets. pp 289--299,
  \mn@doi{10.1017/S1743921308026513}

\bibitem[\protect\citeauthoryear{{Chaplin}, {Elsworth}, {Miller}, {Verner}  \&
  {New}}{{Chaplin} et~al.}{2007}]{2007ApJ...659.1749C}
{Chaplin} W.~J.,  {Elsworth} Y.,  {Miller} B.~A.,  {Verner} G.~A.,   {New} R.,
  2007, \mn@doi [\apj] {10.1086/512543}, \href
  {https://ui.adsabs.harvard.edu/abs/2007ApJ...659.1749C} {659, 1749}

\bibitem[\protect\citeauthoryear{{Christensen-Dalsgaard}
  et~al.,}{{Christensen-Dalsgaard} et~al.}{1996}]{1996Sci...272.1286C}
{Christensen-Dalsgaard} J.,  et~al., 1996, \mn@doi [Science]
  {10.1126/science.272.5266.1286}, \href
  {https://ui.adsabs.harvard.edu/abs/1996Sci...272.1286C} {272, 1286}

\bibitem[\protect\citeauthoryear{{Compton}, {Bedding}, {Ball}, {Stello},
  {Huber}, {White}  \& {Kjeldsen}}{{Compton}
  et~al.}{2018}]{2018MNRAS.479.4416C}
{Compton} D.~L.,  {Bedding} T.~R.,  {Ball} W.~H.,  {Stello} D.,  {Huber} D.,
  {White} T.~R.,   {Kjeldsen} H.,  2018, \mn@doi [\mnras]
  {10.1093/mnras/sty1632}, \href
  {https://ui.adsabs.harvard.edu/abs/2018MNRAS.479.4416C} {479, 4416}

\bibitem[\protect\citeauthoryear{{Cunha} et~al.,}{{Cunha}
  et~al.}{2021}]{2021MNRAS.508.5864C}
{Cunha} M.~S.,  et~al., 2021, \mn@doi [\mnras] {10.1093/mnras/stab2886}, \href
  {https://ui.adsabs.harvard.edu/abs/2021MNRAS.508.5864C} {508, 5864}

\bibitem[\protect\citeauthoryear{{Davies} et~al.,}{{Davies}
  et~al.}{2016}]{2016MNRAS.456.2183D}
{Davies} G.~R.,  et~al., 2016, \mn@doi [\mnras] {10.1093/mnras/stv2593}, \href
  {https://ui.adsabs.harvard.edu/abs/2016MNRAS.456.2183D} {456, 2183}

\bibitem[\protect\citeauthoryear{Foreman-Mackey}{Foreman-Mackey}{2016}]{corner}
Foreman-Mackey D.,  2016, \mn@doi [The Journal of Open Source Software]
  {10.21105/joss.00024}, 1, 24

\bibitem[\protect\citeauthoryear{{Ge}, {Bi}, {Li}, {Liu}, {Tian}, {Yang}, {Liu}
   \& {Yu}}{{Ge} et~al.}{2015}]{2015MNRAS.447..680G}
{Ge} Z.~S.,  {Bi} S.~L.,  {Li} T.~D.,  {Liu} K.,  {Tian} Z.~J.,  {Yang} W.~M.,
  {Liu} Z.~E.,   {Yu} J.,  2015, \mn@doi [\mnras] {10.1093/mnras/stu2391},
  \href {http://adsabs.harvard.edu/abs/2015MNRAS.447..680G} {447, 680}

\bibitem[\protect\citeauthoryear{{Gent} et~al.,}{{Gent}
  et~al.}{2021}]{2021arXiv211106666G}
{Gent} M.~R.,  et~al., 2021, arXiv e-prints, \href
  {https://ui.adsabs.harvard.edu/abs/2021arXiv211106666G} {p. arXiv:2111.06666}

\bibitem[\protect\citeauthoryear{{Gough}}{{Gough}}{1990}]{1990LNP...367.....O}
{Gough} D.~O.,  1990, {in Osaki, Yoji and Shibahashi, Hiromoto, Progress of
  Seismology of the Sun and Stars}.
 Vol. 367, \mn@doi{10.1007/3-540-53091-6, }

\bibitem[\protect\citeauthoryear{{Houdayer}, {Reese}, {Goupil}  \&
  {Lebreton}}{{Houdayer} et~al.}{2021}]{2021A&A...655A..85H}
{Houdayer} P.~S.,  {Reese} D.~R.,  {Goupil} M.-J.,   {Lebreton} Y.,  2021,
  \mn@doi [\aap] {10.1051/0004-6361/202141711}, \href
  {https://ui.adsabs.harvard.edu/abs/2021A&A...655A..85H} {655, A85}

\bibitem[\protect\citeauthoryear{{Houdek} \& {Gough}}{{Houdek} \&
  {Gough}}{2007}]{2007MNRAS.375..861H}
{Houdek} G.,  {Gough} D.~O.,  2007, \mn@doi [\mnras]
  {10.1111/j.1365-2966.2006.11325.x}, \href
  {https://ui.adsabs.harvard.edu/abs/2007MNRAS.375..861H} {375, 861}

\bibitem[\protect\citeauthoryear{{Howe}, {Basu}, {Davies}, {Ball}, {Chaplin},
  {Elsworth}  \& {Komm}}{{Howe} et~al.}{2017}]{2017MNRAS.464.4777H}
{Howe} R.,  {Basu} S.,  {Davies} G.~R.,  {Ball} W.~H.,  {Chaplin} W.~J.,
  {Elsworth} Y.,   {Komm} R.,  2017, \mn@doi [\mnras] {10.1093/mnras/stw2668},
  \href {https://ui.adsabs.harvard.edu/abs/2017MNRAS.464.4777H} {464, 4777}

\bibitem[\protect\citeauthoryear{{Kiefer}, {Schad}, {Davies}  \&
  {Roth}}{{Kiefer} et~al.}{2017}]{2017A&A...598A..77K}
{Kiefer} R.,  {Schad} A.,  {Davies} G.,   {Roth} M.,  2017, \mn@doi [\aap]
  {10.1051/0004-6361/201628469}, \href
  {https://ui.adsabs.harvard.edu/#abs/2017A&A...598A..77K} {598, A77}

\bibitem[\protect\citeauthoryear{{Kjeldsen} \& {Bedding}}{{Kjeldsen} \&
  {Bedding}}{1995}]{1995A&A...293...87K}
{Kjeldsen} H.,  {Bedding} T.~R.,  1995, \aap, \href
  {https://ui.adsabs.harvard.edu/#abs/1995A&A...293...87K} {293, 87}

\bibitem[\protect\citeauthoryear{{Kjeldsen}, {Bedding}  \&
  {Christensen-Dalsgaard}}{{Kjeldsen} et~al.}{2008}]{2008ApJ...683L.175K}
{Kjeldsen} H.,  {Bedding} T.~R.,   {Christensen-Dalsgaard} J.,  2008, \mn@doi
  [\apj] {10.1086/591667}, \href
  {https://ui.adsabs.harvard.edu/#abs/2008ApJ...683L.175K} {683, L175}

\bibitem[\protect\citeauthoryear{{Li}, {Bedding}, {Li}, {Bi}, {Stello}, {Zhou}
  \& {White}}{{Li} et~al.}{2020a}]{2020MNRAS.495.2363L}
{Li} Y.,  {Bedding} T.~R.,  {Li} T.,  {Bi} S.,  {Stello} D.,  {Zhou} Y.,
  {White} T.~R.,  2020a, \mn@doi [\mnras] {10.1093/mnras/staa1335}, \href
  {https://ui.adsabs.harvard.edu/abs/2020MNRAS.495.2363L} {495, 2363}

\bibitem[\protect\citeauthoryear{{Li}, {Bedding}, {Christensen-Dalsgaard},
  {Stello}, {Li}  \& {Keen}}{{Li} et~al.}{2020b}]{2020MNRAS.495.3431L}
{Li} T.,  {Bedding} T.~R.,  {Christensen-Dalsgaard} J.,  {Stello} D.,  {Li} Y.,
    {Keen} M.~A.,  2020b, \mn@doi [\mnras] {10.1093/mnras/staa1350}, \href
  {https://ui.adsabs.harvard.edu/abs/2020MNRAS.495.3431L} {495, 3431}

\bibitem[\protect\citeauthoryear{{Li}, {Li}, {Bi}, {Bedding}, {Davies}  \&
  {Du}}{{Li} et~al.}{2022}]{2022ApJ...927..167L}
{Li} T.,  {Li} Y.,  {Bi} S.,  {Bedding} T.~R.,  {Davies} G.,   {Du} M.,  2022,
  \mn@doi [\apj] {10.3847/1538-4357/ac4fbf}, \href
  {https://ui.adsabs.harvard.edu/abs/2022ApJ...927..167L} {927, 167}

\bibitem[\protect\citeauthoryear{{Lund} et~al.,}{{Lund}
  et~al.}{2017}]{2017ApJ...835..172L}
{Lund} M.~N.,  et~al., 2017, \mn@doi [\apj] {10.3847/1538-4357/835/2/172},
  \href {https://ui.adsabs.harvard.edu/abs/2017ApJ...835..172L} {835, 172}

\bibitem[\protect\citeauthoryear{{Lyttle} et~al.,}{{Lyttle}
  et~al.}{2021}]{2021MNRAS.505.2427L}
{Lyttle} A.~J.,  et~al., 2021, \mn@doi [\mnras] {10.1093/mnras/stab1368}, \href
  {https://ui.adsabs.harvard.edu/abs/2021MNRAS.505.2427L} {505, 2427}

\bibitem[\protect\citeauthoryear{{Paxton}, {Bildsten}, {Dotter}, {Herwig},
  {Lesaffre}  \& {Timmes}}{{Paxton} et~al.}{2011}]{2011ApJS..192....3P}
{Paxton} B.,  {Bildsten} L.,  {Dotter} A.,  {Herwig} F.,  {Lesaffre} P.,
  {Timmes} F.,  2011, \mn@doi [The Astrophysical Journal Supplement Series]
  {10.1088/0067-0049/192/1/3}, \href
  {https://ui.adsabs.harvard.edu/#abs/2011ApJS..192....3P} {192, 3}

\bibitem[\protect\citeauthoryear{{Paxton} et~al.,}{{Paxton}
  et~al.}{2015}]{2015ApJS..220...15P}
{Paxton} B.,  et~al., 2015, \mn@doi [The Astrophysical Journal Supplement
  Series] {10.1088/0067-0049/220/1/15}, \href
  {https://ui.adsabs.harvard.edu/#abs/2015ApJS..220...15P} {220, 15}

\bibitem[\protect\citeauthoryear{{Rasmussen} \& {Williams}}{{Rasmussen} \&
  {Williams}}{2006}]{2006gpml.book.....R}
{Rasmussen} C.~E.,  {Williams} C. K.~I.,  2006, {Gaussian Processes for Machine
  Learning}

\bibitem[\protect\citeauthoryear{{Rendle} et~al.,}{{Rendle}
  et~al.}{2019}]{2019MNRAS.484..771R}
{Rendle} B.~M.,  et~al., 2019, \mn@doi [\mnras] {10.1093/mnras/stz031}, \href
  {https://ui.adsabs.harvard.edu/abs/2019MNRAS.484..771R} {484, 771}

\bibitem[\protect\citeauthoryear{{Salabert}, {R{\'e}gulo}, {P{\'e}rez
  Hern{\'a}ndez}  \& {Garc{\'\i}a}}{{Salabert}
  et~al.}{2018}]{2018A&A...611A..84S}
{Salabert} D.,  {R{\'e}gulo} C.,  {P{\'e}rez Hern{\'a}ndez} F.,   {Garc{\'\i}a}
  R.~A.,  2018, \mn@doi [\aap] {10.1051/0004-6361/201731714}, \href
  {https://ui.adsabs.harvard.edu/#abs/2018A&A...611A..84S} {611, A84}

\bibitem[\protect\citeauthoryear{{Silva Aguirre} et~al.,}{{Silva Aguirre}
  et~al.}{2017}]{2017ApJ...835..173S}
{Silva Aguirre} V.,  et~al., 2017, \mn@doi [\apj]
  {10.3847/1538-4357/835/2/173}, \href
  {https://ui.adsabs.harvard.edu/#abs/2017ApJ...835..173S} {835, 173}

\bibitem[\protect\citeauthoryear{{Sonoi}, {Samadi}, {Belkacem}, {Ludwig},
  {Caffau}  \& {Mosser}}{{Sonoi} et~al.}{2015}]{2015A&A...583A.112S}
{Sonoi} T.,  {Samadi} R.,  {Belkacem} K.,  {Ludwig} H.~G.,  {Caffau} E.,
  {Mosser} B.,  2015, \mn@doi [\aap] {10.1051/0004-6361/201526838}, \href
  {https://ui.adsabs.harvard.edu/abs/2015A&A...583A.112S} {583, A112}

\bibitem[\protect\citeauthoryear{{Townsend} \& {Teitler}}{{Townsend} \&
  {Teitler}}{2013}]{2013MNRAS.435.3406T}
{Townsend} R.~H.~D.,  {Teitler} S.~A.,  2013, \mn@doi [\mnras]
  {10.1093/mnras/stt1533}, \href
  {https://ui.adsabs.harvard.edu/#abs/2013MNRAS.435.3406T} {435, 3406}

\bibitem[\protect\citeauthoryear{{Verma}, {Raodeo}, {Basu}, {Silva Aguirre},
  {Mazumdar}, {Mosumgaard}, {Lund}  \& {Ranadive}}{{Verma}
  et~al.}{2019}]{2019MNRAS.483.4678V}
{Verma} K.,  {Raodeo} K.,  {Basu} S.,  {Silva Aguirre} V.,  {Mazumdar} A.,
  {Mosumgaard} J.~R.,  {Lund} M.~N.,   {Ranadive} P.,  2019, \mn@doi [\mnras]
  {10.1093/mnras/sty3374}, \href
  {https://ui.adsabs.harvard.edu/abs/2019MNRAS.483.4678V} {483, 4678}

\makeatother
\end{thebibliography}



\appendix
\onecolumn
\section{}

\subsection{Inspecting frequency differences between the grid points}

There are two free parameters in the RBF kernel, i.e., the lengthscale and the variance. Proper choices of kernel parameters are the key to the noise model. 
We inspect the frequency change between consecutive grid points to determine the kernel parameters. We present frequency differences between models with approximately the solar mean density in Figure~\ref{fig:app1}. As shown, frequency differences can be described as a smooth function of the frequency plus a damped sine function (the signature of the helium glitch). The systematic uncertainty hence contains two kernels. The primary kernel should have a large lengthscale and a frequency-dependent variance for each input {\bf dimension}. The secondary kernel represents the signature of the helium glitch. It hence has a relatively small lengthscale and some variances decreasing with the frequency.

\begin{figure}
       \center
	\includegraphics[width=0.8\columnwidth]{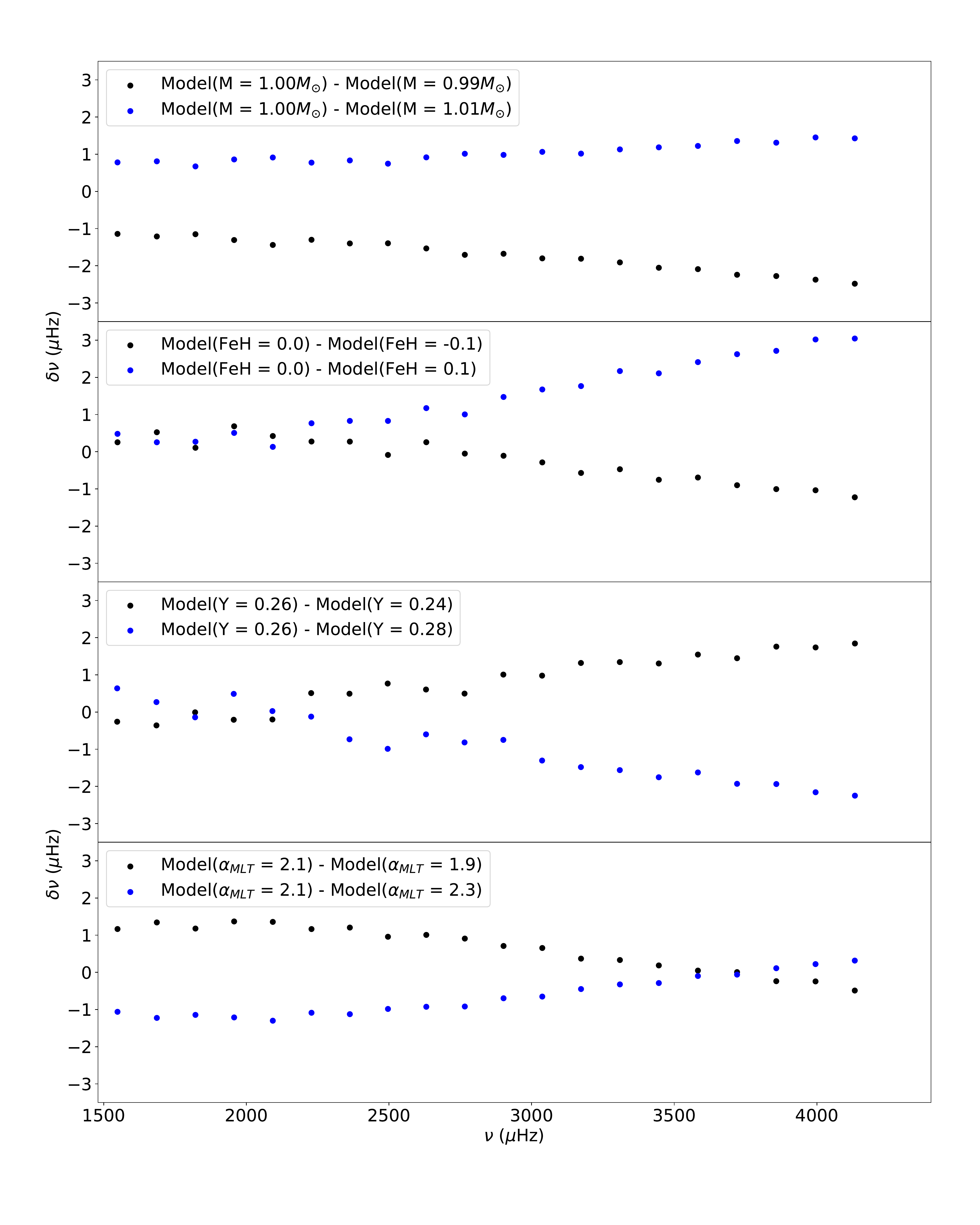}
    \caption{Mode frequency changes between consecutive grid points in four input {\bf dimensions}. All presented models have approximately solar mean density. } 
  \label{fig:app1}
\end{figure}


\bsp	
\label{lastpage}
\end{document}